\documentclass[11pt]{article}

\usepackage[preprint]{acl}

\usepackage{times}
\usepackage{latexsym}
\usepackage[T1]{fontenc}
\usepackage[utf8]{inputenc}
\usepackage{microtype}
\usepackage{inconsolata}
\usepackage{graphicx}
\usepackage{hyperref}

\usepackage{amsmath}
\usepackage{amssymb}
\usepackage{booktabs}
\usepackage{listings} 
\usepackage{subcaption}
\usepackage{adjustbox}

\title{LEAF: Knowledge Distillation of Text Embedding Models with Teacher-Aligned Representations%
\thanks{To appear in \textit{Proceedings of the 64th Annual Meeting of the Association for Computational Linguistics (ACL 2026)}.}}

\author{Robin Vujanic\thanks{Corresponding author.} \and Thomas Rückstieß \\
MongoDB Research \\
\texttt{robin.vujanic@mongodb.com, me@tomr.au}\\
}

\begin{document}
  \newcommand{\robincom}[1]{
  {
      \color{blue}
      \begin{flushright}
        \framebox{
          \parbox[t]{0.95\linewidth}{
            \underline{Comment RV}
            \par
            \em #1
          }
        }
      \end{flushright}
    }
}
\newcommand{\blue}[1]{\color{blue}#1\color{black}}
\newcommand{\red}[1]{\color{red}#1\color{black}}
\newcommand{\gray}[1]{\color{gray}#1\color{black}}

\newcommand{\ub}[1]{\underline{\textbf{#1}}}

\newcommand{\R}{\mathbb{R}}
\newcommand{\N}{\mathbb{N}}

\newcommand{\adept}{\texttt{leaf}}
\newcommand{\LEAF}{\texttt{LEAF}}
\newcommand{\leaf}{\texttt{leaf}}
\newcommand{\adeptir}{\hbox{\texttt{leaf-ir}}}
\newcommand{\adeptmt}{\hbox{\texttt{leaf-mt}}}

\newcommand{\efivelarge}{\hbox{\texttt{e5-large}}}

\newcommand{\vocab}{Vocabulary}
\newcommand{\snowflakem}{arctic-embed-m-v1.5}
\newcommand{\snowflakexs}{arctic-embed-xs}
\newcommand{\openaismall}{OpenAI 3 (small)}
\newcommand{\openaismallname}{\texttt{text-embedding-3-small}}
\newcommand{\mxbai}{mxbai-l-v1}
\newcommand{\minilm}{MiniLM-L6-v2}

\newcommand{\wout}{W^\mathrm{out}}
\newcommand{\dimembed}{d}
\newcommand{\dimadept}{d'}
\newcommand{\Ll}{L}
\newcommand{\Lp}{L'}

\newcommand{\tr}{^\intercal}
\newcommand{\yhat}{\hat{y}}

  \maketitle

  \begin{abstract}
    We present \LEAF{} (``\textbf{L}ightweight \textbf{E}mbedding \textbf{A}lignment \textbf{F}ramework''), a knowledge distillation framework for text embedding models.
    A key distinguishing feature is that our distilled \adept{} models are compatible with their teacher,
    enabling
    flexible asymmetric architectures where documents are encoded
    with the larger teacher model, while queries use smaller \adept{} models.
    We also show that
    \adept{} models automatically inherit MRL and robustness to output quantization whenever these properties are
    present in the teacher model, without explicitly training for them.
    To demonstrate the effectiveness of our framework we publish \adeptir, a 23M parameters information retrieval oriented model
    focusing on English texts
    that, besides being teacher-compatible, sets a new state-of-the-art (SOTA) on BEIR, ranking no.1 on the public
    leaderboard for models of its size.
    Asymmetric mode further increases its retrieval performance.
    Our scheme is however not restricted to information retrieval. We demonstrate its wider
    applicability by synthesizing the multi-task \adeptmt{} model.
    This also sets a new SOTA, achieving no.1
    on the public MTEB v2 (English) leaderboard for models of its size.
    \LEAF{} is applicable to black-box models,
    requires no judgments nor hard negatives,
    and training can be conducted using small batch sizes.
    Thus, dataset and training infrastructure requirements
    for our framework are modest. We make our models publicly available under a permissive Apache~2.0 license.
  \end{abstract}

  \begin{figure*}[t]
    \centering
    \includegraphics[width=\linewidth]{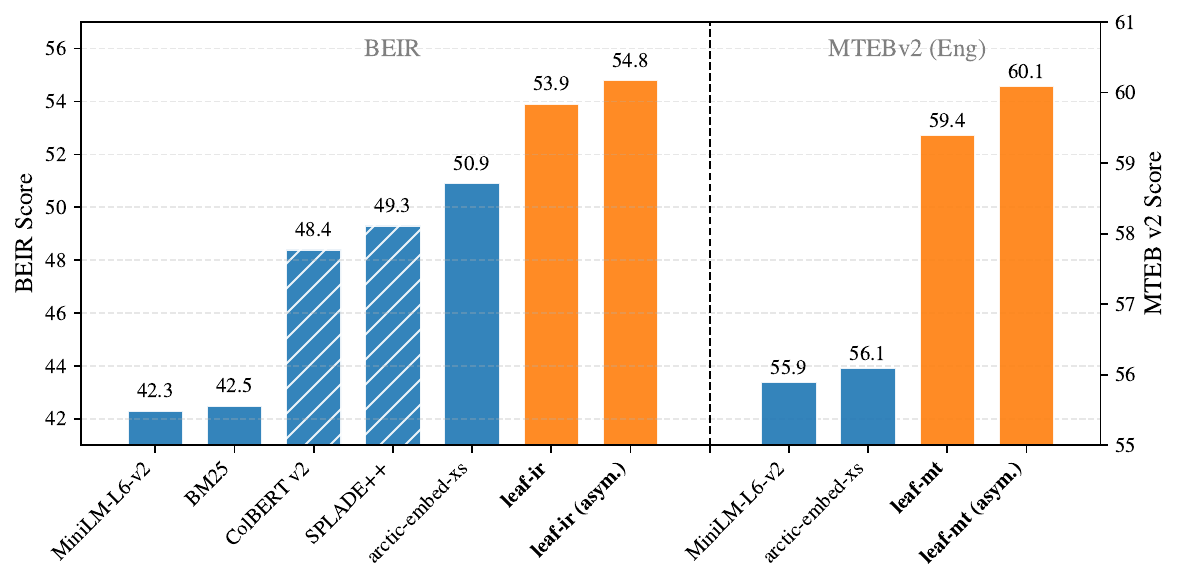}
    \caption{Our information retrieval oriented \adeptir{} model (23M parameters) sets a new state-of-the-art on the BEIR benchmark for $\leq$100M parameters models. When run in asymmetric mode, its retrieval performance is
    further increased. Our multi-task \adeptmt{} model (23M parameters) also sets a new state-of-the-art on MTEB v2 (English). Hatched columns indicate comparison models that leverage larger (110M parameters) models.}
    \label{fig:teaser}
  \end{figure*}

  \section{Introduction}
\label{sec:introduction}

Recent advancements in neural networks have paved the way for drastic improvements in a wide array of natural language processing tasks,
and
new use cases like
retrieval augmented generation (``RAG'') are fueling a massive increase in user demand for these Transformer-based~\cite{vaswani2017attention} models.


Unfortunately, this dramatic performance jump has come at the cost of an explosion in the size of these models,
resulting in massive costs in terms of hardware, electric power, and maintenance required to train and operate them.
To accommodate varying budgets,
model providers typically offer a variety of model sizes.
In the case of bi-encoder text embedding models, however,
none of these models, even different sizes from the same model family, are compatible with each other to the best of our knowledge.
As a result, users desiring to switch to a different embedding model are required to perform a costly complete recomputation of the embeddings of all their data.
In the context of information retrieval (IR), this incompatibility also leads to rigid architectures where the same large model
needs to be utilized to encode both documents as well as queries.
But the system requirements of these two phases of IR are not symmetric:
while documents are typically embedded only once at index time and under mild
latency requirements, embedding user queries using large models can cause substantial sustained operational costs and can incur prohibitive latencies in end-user experience.

\begin{figure}[t]
  \centering
  \includegraphics[width=1.0\linewidth]{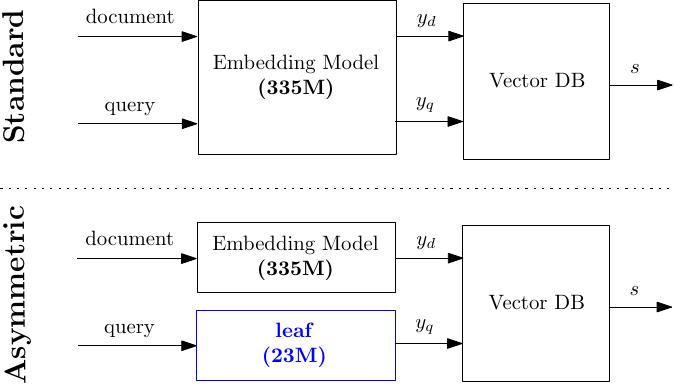}
  \caption{\LEAF-enabled asymmetric architecture.}
  \label{fig:new_paradigm}
\end{figure}

To address these challenges we propose \LEAF, a knowledge distillation \emph{framework} consisting of model architecture, training loss,
training regime and training datasets. \LEAF{} produces text embedding models that are \emph{aligned} to their teacher.
The compatibility of \leaf{} models enables flexible architectures like the one shown in Figure~\ref{fig:new_paradigm}
where document embeddings are computed using large and expensive models, while queries can be more economically encoded with
the smaller \leaf{} models.

We demonstrate the efficacy of our framework by first synthesizing an information retrieval oriented model focused on English texts called \adeptir.
This 23M parameter model is a distillation of a 109M teacher (4.7$\times$ compression), leading to a 6.5$\times$ and $7.3\times$ throughput increase when
encoding documents and queries respectively.
When run in standard mode, encoding both queries and documents, \adeptir{} sets a new SOTA BEIR~\cite{beir} score for
models of its size, ranking no.1 on the corresponding public leaderboard at the time of writing.


One of our core contributions is to show that taking advantage of the asymmetric system
in Figure~\ref{fig:new_paradigm} unlocks further effectiveness gains.
Namely, Figure~\ref{fig:teaser} illustrates that we set even higher SOTA scores in settings where query
time has sufficient computational budget only to run the student model, while documents can be processed
with the larger teacher.
We generally find that IR performance of the asymmetric system is somewhere in the middle between running the
teacher and the student models in their respective standard modes.

Our scheme is not specialized to IR\@.
To demonstrate \LEAF's broader applicability we synthesize the multi-task
\adeptmt{} model,
by performing an even more aggressive distillation of 335M parameters down to 23M\@.
This leads to throughput gains by a factor of 24.4$\times$ and 23.7$\times$ for documents and queries respectively.
We assess its performance on MTEB v2-English~\citep{mteb_eng_v2}, a benchmark that, besides IR, tests models'
performance on other tasks such as classification, clustering, reranking, semantic sentence similarity, and summarization.
When run in standard mode, \adeptmt{} also sets a new SOTA
on this benchmark and ranks no.1 on the corresponding public leaderboard for models of its size.
Taking advantage of asymmetric mode further increases performance also in this case, as shown on Figure~\ref{fig:teaser}.

We furthermore show that \adept{}models automatically inherit Matryoshka Representation Learning
(MRL, ~\citealp{mrl}) and robustness to quantization properties whenever the teacher model has them,
without specifically training for them.
MRL allows the truncation of embedding vectors at arbitrary lengths, while vector quantization allows the storage of
embedding vectors using more compact types, i.e., \verb|int8| or \verb|binary| instead of \verb|float32|.
Both of these techniques aim at reducing storage space and increasing retrieval speed, by gradually trading off retrieval performance.

\LEAF{} can be applied to black-box models since, in contrast to several other knowledge distillation frameworks~\cite{minilm,distilbert,tinybert,mobilebert}, it
does not require access to any of the model's internals such as keys, queries and values.
Further, it can be applied in cases where the architectures of \leaf{} and its teacher are different: they can,
for instance, use different tokenizers, have a different number of layers and attention heads, or different hidden state dimensions.


A second advantage of \LEAF{} is that it is not based on contrastive loss and hence does not require the availability
of judgments or hard negatives as training datasets.
These are typically needed to train embedding models~\cite{openai_embed,splade_v1,colbert_v2}
as well as some other knowledge distillation procedures designed for text embedding
models~\cite{splade_v2_distillation}.

Third, as elaborated in Section~\ref{sec:model-training}, training works well also with small batch sizes.
As a result, we were able to train the two aforementioned SOTA models on a single A100 GPU within a budget of 100 hours each.

We describe \LEAF{} as a \emph{lightweight} knowledge distillation scheme, thanks to the deliberate simplicity of the loss,
entailing exclusively the $\ell_2$ norm of the approximation error between student and teacher representations,
and model architecture, along with simplified training dataset and training infrastructure requirements.
A key contribution of our work is to show that such an uncomplicated setup is sufficient to derive SOTA text
embedding models.


Finally, we reflect on the system's robustness to perturbations.
For instance, we expect documents with similar meanings
but different formulations to score similarly against a given query.
We consider our work as seeking to synthesize models that directly approximate the
function that maps strings to embeddings of their teacher.
In Section~\ref{sec:systems_robustness} we find that downstream task performance is
similarly robust to the perturbations introduced by our approximation scheme.
We observe that a substantial amount of approximation error can be absorbed by
the system and call this its \emph{robustness margin}.

\section{Related Work}
\label{sec:related-work}

Despite the simplicity of the approach proposed in this paper, literature explicitly exploring its potential
remains sparse.

Closest to our work is~\cite{kale_model}, which studies asymmetric retrieval supported by
query and document encoders of different sizes. Distillation of the query encoder is
achieved via layer pruning of a Transformer backbone pre-trained to be compatible with the
document encoder, followed by alignment using a KL divergence loss. However, this work does
not aim to synthesize competitive embedding models, but rather to analyze the impact of pruning
on retrieval performance. Further, we allow independently trained query and document encoders
and do not assume access to teacher weights.

The work in~\cite{embedding-converter}, although not aimed at solving the query/document asymmetry problem,
trains an MLP converter network to translate embeddings between models. Unlike our findings, they
report that regression alone is insufficient and consequently introduce additional loss components.
In their setup, both Transformer models are frozen and only the MLP is trained, which we believe explains the differing
conclusions.

The recipe in~\cite{sentencetransformers_distillation} down-scales teacher embeddings to the
student’s dimensionality using PCA and aligns them via MSE loss. However, PCA destroys properties
such as MRL and robustness to quantization. Moreover, since we propose to use the teacher for document
encoding and optionally a student for queries, dimensionality reduction degrades
document representations regardless of whether the student is deployed. Our approach avoids this issue by training
students directly in the teacher’s embedding space.

MSE loss has also been used in~\cite{reimers_making_monolingual_models_multilingual} to extend monolingual
models to multilingual settings, although in this case the student is typically larger than the teacher.

Several prior works propose distillation schemes for Transformer-based language models,
including TinyBERT~\cite{tinybert,tinybert_simplified}, DistilBERT~\cite{distilbert},
and MobileBERT~\cite{mobilebert}. These methods target language modeling heads and are not
directly applicable to embedding models. MiniLM~\cite{minilm} is applicable to embeddings but
requires access to model internals, matching tokenizers, and the same number of attention heads.
\cite{emo_distillation} proposes a solution to the problem of distilling models
with different tokenizers using optimal transport ideas.

Most commonly, embedding models are distilled using similarity scores.
The approach in~\cite{splade_v2_distillation} introduces a Margin MSE loss, later adopted by SPLADE
v2~\cite{splade_v1}, but relies on judgments and hard negatives, unlike our method.
KL loss on the softmax of similarity scores of query and documents is applied in~\cite{qwen3_vl_embedding_and_reranker}
to distill from a reranker teacher, while~\cite{m3_paper} performs self-distillation taking advantage of the
fact that the same model can be used to produce dense as well as sparse representations.

\section{Approach}
\label{sec:approach}

\begin{figure}[h]
  \centering
  \includegraphics[width=\linewidth]{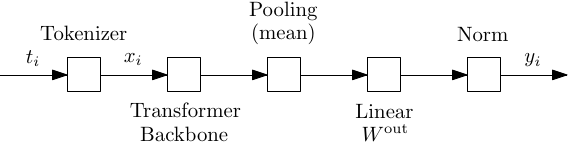}
  \caption{\leaf{} model architecture.}
  \label{fig:model_architecture}
\end{figure}

The architecture used in this work is
shown in Figure~\ref{fig:model_architecture}.
It consists of a Transformer $\mathrm{Backbone}$,
followed by mean pooling. In order to match the target teacher model's output dimension $d$ we stack $\wout \in \R^{\dimadept \times \dimembed}$,
a $\mathrm{Linear}$ layer that maps the Backbone's (typically lower) dimensions $\dimadept$ into $\dimembed$.


Let $I$ be an index over the training set, and
\begin{math}
  \left\{ (t_i, \yhat_i) \right\}_{i \in I}
\end{math}
be the collection of tuples of training texts $t_i$ (strings) and their corresponding ground truth embeddings $\yhat_i \in \R^d$
produced by the teacher model.
The texts $t_i$ can entail both queries as well as documents. Let
\begin{math}
  x_i \in \mathbb{N}^T
\end{math}
be the tokenization of $t_i$, where $T$ is the sequence length. Padding
tokens \verb|[PAD]| are appended to make the lengths of sequences in a batch match. Respectively, sequences are
truncated if they exceed the model's maximum context length $\overline{T}$.

Then, the embedding vector $y_i \in \R^{d}$ computed by our model for a given input sequence $x_i$ is given by
\begin{equation}
  \label{eq:adept_model}
  y_i = \mathrm{Norm}(\mathrm{Linear}(\mathrm{Mean}(\mathrm{Transf.}(x_i)))),
\end{equation}
where mean pooling $\mathrm{Mean}$ is computed only over Transformer outputs corresponding to non-\verb|[PAD]| tokens.
We use mean pooling independently of the type of pooling utilized by the teacher model.
We have experimented with other popular pooling methods implemented in
\texttt{sentence\_transformers}\footnote{\url{https://sbert.net/}},
namely \texttt{[CLS]}- and \texttt{[EOS]}-token pooling, as well as max pooling, but found them to be inferior choices,
see Appendix~\ref{appendix:pooling_layer_ablation}.
The normalization layer $\mathrm{Norm}(\cdot)$ is added to the stack if the teacher model
is structured to produce normalized embedding vectors, i.e.,
\begin{equation*}
  \|\yhat_i\|_2 = 1 \quad \forall i \in I.
\end{equation*}

In this work, we intend to train the model given by Eq.~\eqref{eq:adept_model} so that $y_i$ approximates $\yhat_i$.
Accordingly, we define the approximation error
\begin{equation}
  e_i = y_i - \yhat_i,
\end{equation}
and utilize it as a knowledge distillation training loss:
\begin{equation}
  \label{eq:training_loss}
  \mathcal{L}_i = \| e_i \|_2.
\end{equation}
We propose to limit the training loss to the $\ell_2$ expression in Eq.~\eqref{eq:training_loss}.
This choice enables the distillation procedure presented in this paper to be applicable also in
cases when no internals of the models, such as keys, queries and values, are known.
That is, it can be applied to any black-box model for which the training tuples
\begin{math}
  (t_i, \yhat_i)
\end{math}
can be obtained. Note also that, as mentioned in the Introduction, the computation of this loss does not require
judgments nor hard negatives, and that it is also not specific to the information retrieval (IR) task.

Appendix~\ref{appendix:kd_ablation} reports training loss alternatives to Eq.~\eqref{eq:training_loss} based
on other knowledge distillation approaches available in the literature. Specifically, we extended Eq.~\eqref{eq:training_loss}
with the losses proposed by
TinyBERT \cite{tinybert}, DistilBERT \cite{distilbert} and MiniLM \cite{minilm}. These, however,
did not produce better results in our experiments, and did not have all the benefits enumerated above,
so we did not pursue them further. It should also be noted that, as shown in the Appendix,
these distillation procedures require the tokenizers to match, and in
some cases they also require additional internals, such as the number of attention heads in the Transformer
layers, to be the same. Our procedure does not have these restrictions.




\subsection{Training Datasets}
\label{sec:training-datasets}

We selected our training datasets to cover a broad range of topics, including general knowledge, news, science, entertainment,
and commerce. Among them, \vocab~ is a new dataset that we created in the context of this work and that we are making publicly
available. \vocab~ was synthesized by taking a list of 479k words appearing in English language contexts from \cite{500k_words}
and prompting Claude 3.5 Sonnet\footnote{\url{https://www.anthropic.com/}} to produce definitions or important facts about them.
Appendix~\ref{appendix:vocab_dataset} reports the prompt used as well as samples from this dataset.

All the other datasets are also publicly available. We do not perform any processing of the raw data contained in them;
we merely extract the relevant column and store it as a parquet file in our processing pipelines.

Our training data consists of both document and query segments. For documents, we utilize 3M samples from FineWeb \cite{fineweb}
(sampled from the 10B tokens subsample), 900k samples from CC-News (100k random documents for
each year between 2016 and 2024), 30k BERT tokens (representing each individual token
from BERT's tokenizer~\citealp{bert_devlin} to allow for baseline alignment), and 800k
entries from our \vocab~ dataset. For queries, we use 979k samples from Amazon QA~\cite{amazonqa},
27k from \verb|LoTTE/pooled/| \cite{colbert_v2}, 502k from \verb|MSMARCO/train/| \cite{msmarco},
272k from \verb|PubMedQA/pqa_artificial/| and \verb|PubMedQA/pqa_unlabeled/| \cite{pubmedqa},
and 73k from Trivia QA~\cite{triviaqa}.

We stress the fact that although some of these datasets also entail judgments, we do not utilize them for training since they are not needed in the computation of the loss in Eq.~\eqref{eq:training_loss}.

\begin{figure}[hbtp]
  \caption{Model performance when trained for one epoch with different sub-segments of training datasets.}
  \label{fig:training_dataset_ablation}
  \includegraphics[width=1\linewidth]{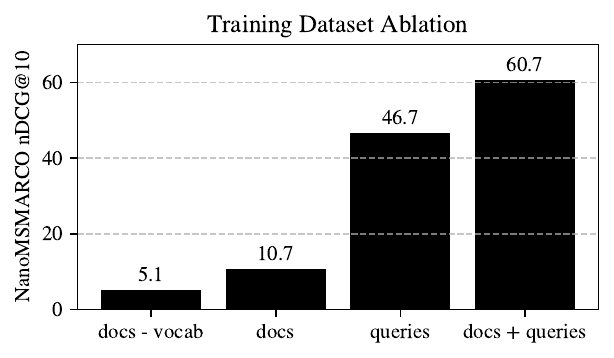}
\end{figure}

Figure~\ref{fig:training_dataset_ablation} shows the relative impact of these training datasets on downstream performance of \adeptir{} on NanoMSMARCO, after training for 1 epoch. These results indicate that training on queries is more important than documents, but to reach state-of-the-art performance it is necessary to train on both.

We finally note that we have reviewed the sources of each one of these datasets to avoid contamination with the datasets
utilized for the benchmarking results discussed in Section~\ref{sec:results}. In particular, we consider MSMARCO and its smaller variant NanoMSMARCO as
training/dev
datasets, as is common in the literature, and do not include them in any of our performance benchmarks.

\subsection{Model Training}
\label{sec:model-training}

The Transformer backbone used in this work is \verb|MiniLM-L6-v2| \cite{minilm}, a 23M parameter encoder model.
We train the parameters of the whole model, which includes the Transformer backbone and $\wout$.
We use AdamW as the optimizer, setting the initial learning rate to 1e-4 and leave the other parameters at their
defaults ($\beta_1=0.9$, $\beta_2=0.999$, weight decay $=0.01$).
We train our model for 3 cycles of linear learning rate decay over 10 epochs, resulting in a total of 30 training epochs.
The learning rate is linearly decayed to 1e-5 over the 10 epochs of each cycle and then reset to 1e-4
before the following cycle.
Appendix~\ref{appendix:learn_rate_decay_schedule} reports our experimentation
comparing different learning rate schedules. We found linear decay to work best.




We set the batch size to 32. As shown in \cite{contrastive_batch_sizes_simclr}, approaches that rely on contrastive loss
typically benefit from larger batch sizes. For instance, \cite{grit} uses a batch size of 2,048
while in \cite{arctic_embed_v2} the batch size is set to 32,768.
In our work we found that a batch size of 32 with our loss in Eq.~\eqref{eq:training_loss} is sufficient.
In fact, the analysis in Appendix~\ref{appendix:batch_size_selection} shows that our training regime favors more steps with
smaller batch sizes, rather than the other way around. We believe this is owed to the $\ell_2$ loss in Eq.~\eqref{eq:training_loss}
providing dense supervision across all embedding dimensions, unlike contrastive approaches that only provide relative
ordering supervision between positive and negative pairs.

We precompute and cache the target embeddings $\yhat_i$ for all the training samples
in the datasets discussed in Sec.~\ref{sec:training-datasets}. This substantially decreases training time as we only need
to perform forward passes on the student \leaf{} model. Whenever a model supports instruction prompts
\cite{instructor_instruction_prompts_embedding_models}, we supply them before computing the corresponding ground truth
embedding $\yhat_i$, and we also supply them to \adept{} when we compute $y_i$ according to Eq.~\eqref{eq:adept_model}.

Overall, we obtain approximately 200k training batches from the datasets described in
Section~\ref{sec:training-datasets}, out of which we hold out a randomly sampled
validation set of 128 batches.
The remaining training batches are shuffled before each epoch.
With this setup, we train on a single NVIDIA A100 40GB GPU with a budget of 100 hours.



  \section{Results}
\label{sec:results}

\subsection{Information Retrieval Model}
\label{sec:results_ir}

\textbf{\adeptir{} is teacher-compatible and sets a new state-of-the-art for compact information retrieval oriented embedding models, while supporting MRL and vector quantization.}

We trained \adeptir, a model focused on information retrieval (IR). We make this model publicly available under a permissive Apache~2.0 license.

We assess its effectiveness on BEIR~\cite{beir}, an industry standard benchmark for IR systems.
Table~\ref{tab:beir-results-main} reports our results on this benchmark, and Figure~\ref{fig:teaser} displays them graphically.
We compare against \snowflakexs~\cite{arctic_v1}, the current state-of-the-art
for $\leq$30M parameters models, as well as other popular retrieval models, including SPLADE++~\cite{splade_v1} and ColBERT v2 \cite{colbert_v2}, although we note that these leverage larger
(110M parameters) BERT models, and are thus hatched in Figure~\ref{fig:teaser}.
We also include baseline BM25~\cite{bm25_benchmark} scores.
The last row reports the scores of \snowflakem~\cite{arctic_v1},
the teacher model \adeptir{} was distilled from. This teacher model was chosen for its strong IR performance at its size.

\begin{table*}[t]
  \centering
  \small
  \caption{nDCG@10 scores on the BEIR \cite{beir} benchmark for \adeptir. $^\dagger$BM25 scores are obtained with $(k_1=0.9, b=0.4)$. SPLADE++ and ColBERT v2 scores are from \cite{tite_paper}, while scores for \snowflakem, \snowflakexs~, and \minilm~ are from the public MTEB leaderboard.}
  \label{tab:beir-results-main}
  \setlength{\tabcolsep}{4pt}
  \begin{tabular}{l|c|*{14}{c}|c}
    \toprule
    \multicolumn{1}{c}{}       & \multicolumn{1}{c}{\textbf{Size}} & \multicolumn{1}{c}{\adjustbox{angle=45,lap=\width}{\textbf{ArguAna}}} & \multicolumn{1}{c}{\adjustbox{angle=45,lap=\width}{\textbf{ClimateFEVER}}} & \multicolumn{1}{c}{\adjustbox{angle=45,lap=\width}{\textbf{CQADupStack}}} & \multicolumn{1}{c}{\adjustbox{angle=45,lap=\width}{\textbf{DBPedia}}} & \multicolumn{1}{c}{\adjustbox{angle=45,lap=\width}{\textbf{FEVER}}} & \multicolumn{1}{c}{\adjustbox{angle=45,lap=\width}{\textbf{FiQA2018}}} & \multicolumn{1}{c}{\adjustbox{angle=45,lap=\width}{\textbf{HotpotQA}}} & \multicolumn{1}{c}{\adjustbox{angle=45,lap=\width}{\textbf{NFCorpus}}} & \multicolumn{1}{c}{\adjustbox{angle=45,lap=\width}{\textbf{NQ}}} & \multicolumn{1}{c}{\adjustbox{angle=45,lap=\width}{\textbf{Quora}}} & \multicolumn{1}{c}{\adjustbox{angle=45,lap=\width}{\textbf{SCIDOCS}}} & \multicolumn{1}{c}{\adjustbox{angle=45,lap=\width}{\textbf{SciFact}}} & \multicolumn{1}{c}{\adjustbox{angle=45,lap=\width}{\textbf{TREC-COVID}}} & \multicolumn{1}{c}{\adjustbox{angle=45,lap=\width}{\textbf{Touche2020}}} & \multicolumn{1}{c}{\textbf{Avg.}} \\
    \midrule
    \textbf{\adeptir{} (asym.)} & 23M                               & \blue{\ub{59.0}}                                                      & \blue{\ub{37.5}}                                                           & \blue{\ub{42.4}}                                                          & \blue{\ub{45.0}}                                                      & \ub{86.5}                                                           & \blue{\ub{41.3}}                                                       & \blue{68.5}                                                            & \blue{\ub{36.2}}                                                       & \blue{\ub{61.2}}                                                 & 86.0                                                                & \blue{20.3}                                                           & \blue{70.2}                                                           & \blue{\ub{82.6}}                                                         & 30.1                                                                     & \blue{\ub{54.8}}                  \\
    \textbf{\adeptir}          & 23M                               & \ub{58.4}                                                             & \ub{34.6}                                                                  & \ub{42.3}                                                                 & \ub{44.6}                                                             & \ub{86.6}                                                           & \ub{38.4}                                                              & 68.1                                                                   & \ub{35.8}                                                              & \ub{58.9}                                                        & 86.3                                                                & 19.7                                                                  & 70.0                                                                  & \ub{80.3}                                                                & 30.2                                                                     & \ub{53.9}                         \\
    \midrule
    \multicolumn{17}{c}{Comparisons}                                                                                                                                                                                                                                                                                                                                                                                                                                                                                                                                                                                                                                                                                                                                                                                                                                                                                                                                                                                                                                                                                                                            \\
    \midrule
    \snowflakexs               & 23M                               & 52.1                                                                  & 29.9                                                                       & 40.1                                                                      & 40.2                                                                  & 83.4                                                                & 34.5                                                                   & 65.3                                                                   & 30.9                                                                   & 54.8                                                             & 86.6                                                                & 18.4                                                                  & 64.5                                                                  & 79.4                                                                     & 32.8                                                                     & 50.9                              \\
    \minilm                    & 23M                               & 50.2                                                                  & 20.3                                                                       & 41.3                                                                      & 32.3                                                                  & 51.9                                                                & 36.9                                                                   & 46.5                                                                   & 31.6                                                                   & 43.9                                                             & \ub{87.6}                                                           & \ub{21.6}                                                             & 64.5                                                                  & 47.2                                                                     & 16.9                                                                     & 42.3                              \\
    BM25$^\dagger$             & --                                & 40.8                                                                  & 16.2                                                                       & 28.2                                                                      & 31.9                                                                  & 63.8                                                                & 23.8                                                                   & 62.9                                                                   & 31.8                                                                   & 30.5                                                             & 78.7                                                                & 15.0                                                                  & 67.6                                                                  & 58.9                                                                     & \ub{44.2}                                                                & 42.5                              \\
    SPLADE++                   & 110M                              & 52.0                                                                  & 23.0                                                                       & 33.4                                                                      & 43.7                                                                  & 78.8                                                                & 34.7                                                                   & \ub{68.7}                                                              & 34.7                                                                   & 53.8                                                             & 83.4                                                                & 15.9                                                                  & \ub{70.4}                                                             & 72.7                                                                     & 24.7                                                                     & 49.3                              \\
    ColBERT v2                 & 110M                              & 45.3                                                                  & 17.6                                                                       & 35.9                                                                      & 44.1                                                                  & 77.4                                                                & 34.6                                                                   & 66.5                                                                   & 33.0                                                                   & 54.7                                                             & 85.1                                                                & 15.0                                                                  & 69.1                                                                  & 73.2                                                                     & 25.7                                                                     & 48.4                              \\
    \midrule
    \multicolumn{17}{c}{Teacher}                                                                                                                                                                                                                                                                                                                                                                                                                                                                                                                                                                                                                                                                                                                                                                                                                                                                                                                                                                                                                                                                                                                                \\
    \midrule
    \snowflakem                & 109M                              & 59.5                                                                  & 36.9                                                                       & 45.0                                                                      & 45.6                                                                  & 88.4                                                                & 42.4                                                                   & 72.2                                                                   & 36.2                                                                   & 62.5                                                             & 87.4                                                                & 21.5                                                                  & 71.6                                                                  & 84.6                                                                     & 31.4                                                                     & 56.1                              \\
    \bottomrule
  \end{tabular}
\end{table*}

Scores in the table are bold and underlined when our models improve upon the best comparison methods.
We further highlight in blue color scores for which asymmetric mode is an improvement over standard mode.

Our model \adeptir{} ranks no.1 on the public leaderboard\footnote{\url{https://huggingface.co/spaces/mteb/leaderboard}} of BEIR for models with $\leq$100M parameters at the time
of writing, setting a new state-of-the-art for models of this size.
As shown in the table, we set a new state-of-the-art on 9/14 datasets when \adeptir{} is run in standard mode.
Our overall average score also sets a new state-of-the-art at an nDCG@10 of 53.9, retaining 96.1\% of the teacher's IR performance with
$4.7\times$ fewer parameters. On tests run on an AWS EC2 \texttt{i3.large} instance, which is a commonly used
CPU-only VM for database and search workloads, this leads to inference time speed-ups of 6.5$\times$ / 7.3$\times$ for
docs and queries respectively; details are in Appendix~\ref{appendix:inference_time_speed}.



In asymmetric mode, retrieval performance is further increased
in 11/14 datasets compared to \adeptir{} standard,
as highlighted in blue in the table.
It achieves an aggregated score of 54.8 nDCG@10, i.e., it retains 97.7\% of the teacher's performance while running
a $4.7\times$ smaller model at query time.

These results demonstrate that despite its implementation simplicity and low training and infrastructure
requirements, and under a substantial compression regime, \LEAF{} is able to synthesize a general purpose
SOTA retrieval embedding model that is aligned to its teacher.
These results also indicate that \LEAF{} is a better approach to synthesizing smaller variants of a model family
than training via contrastive loss.
In other words, larger models are better suited to optimize the structure of embedding spaces during contrastive training,
while smaller models more easily approximate these pre-optimized structures than build them independently.
This is evidenced by the fact that we substantially outperform \snowflakexs, another 23M parameters model
trained using the same procedure and datasets as \snowflakem, the teacher of \adeptir.

\begin{figure*}[htbp]
  \centering
  \includegraphics[width=\linewidth]{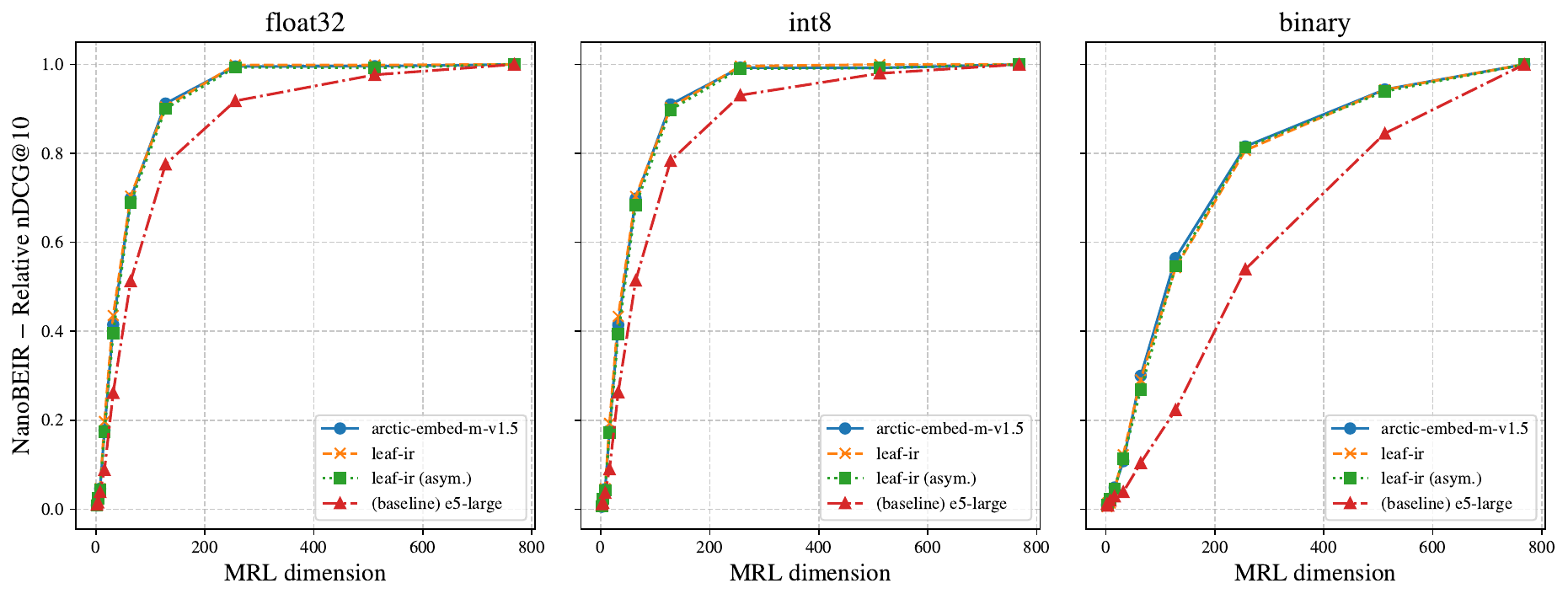}
    \caption{Performance curves for various MRL truncations, as well as output quantization regimes.}
  \label{fig:mrl_transfer-leaf-ir}
\end{figure*}

Figure~\ref{fig:mrl_transfer-leaf-ir} shows that \adeptir{} also inherits MRL \cite{mrl} and vector quantization properties
from the teacher model, without specifically training for them.
The values shown in the figure are relative to the max nDCG\@10 measured for the given model and quantization,
so they vary between 0 and 1.
For comparison, the figure also entails a baseline curve for \efivelarge, a model that has not been trained with MRL.
The figure illustrates that our distillation has an MRL performance profile that is nearly identical to that of its teacher.
Absolute MRL performance scores for \adeptir{} are provided in Appendix~\ref{appendix:mrl_transfer_mt}.

Figure~\ref{fig:tsne_visualization} shows the projection of the embeddings
produced by \adeptir{} and its teacher, for sample texts taken from 7 different domains (sports, technology, finance, \dots).
When projected, the vectors are nearly indistinguishable.
The texts used for this plot are in Appendix~\ref{appendix:geometric_comparison_examples}.
Finally, Appendix~\ref{appendix:scoring_examples} reports an example query over a corpus of 100k documents from CC-News,
comparing the documents retrieved by the teacher and \adeptir{} (in standard and asymmetric modes),
as well as queries and docs from NanoBEIR where approximation errors were highest (and lowest).

\begin{figure}[]
  \centering
  \includegraphics[width=\linewidth]{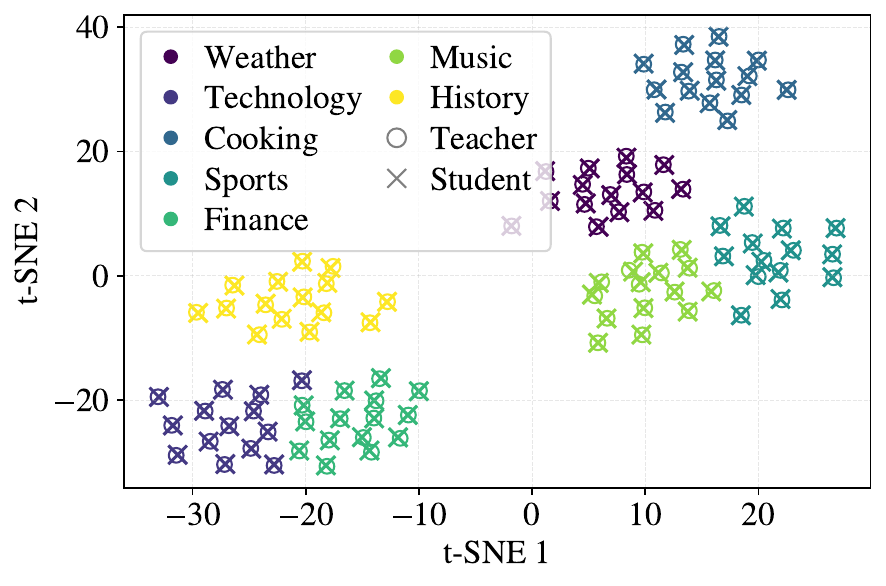}
    \caption{2D projection of embeddings produced by the student and teacher for text covering 7 domains.}
  \label{fig:tsne_visualization}
\end{figure}

\subsection{Multi-Task Model}
\label{sec:results_mt}

\textbf{\adeptmt{} is teacher-compatible and sets a new state-of-the-art for compact multi-task text embedding models, while supporting MRL and vector quantization.}

The training recipe detailed in Section~\ref{sec:approach} is not specific to information retrieval.
We thus trained a 23M parameter multi-task \adeptmt{} model, distilling from \mxbai~ (335M parameters).
The compression ratio in this case is $14.6\times$. This model is also publicly available
under a permissive Apache~2.0 license.

We assess its performance on MTEB v2-English~\cite{mteb_eng_v2}, a multi-task benchmark that,
besides retrieval and reranking, also measures classification, clustering, pair classification, semantic textual similarity, and summarization performance.

Table~\ref{tab:multi-task-results} summarizes the results for this benchmark, and the right hand side of Figure~\ref{fig:teaser}
displays them graphically. Scores for the individual datasets constituting this benchmark can be found in
Appendix~\ref{appendix:mteb_v2_scores_adept_mt}.
We compare against the performance of \minilm~ and \snowflakexs.
At the time of writing, these two models set the previous state-of-the-art for $\leq$30M parameter models\footnote{We have
  excluded from our comparisons models for which the training datasets are not known, although we outperform them too.
}:
\minilm~ achieves the best Borda score, while \snowflakexs~ has the highest average benchmark
score at $56.1$\footnote{MTEB v2 scores are an aggregate, dimensionless score.}.

\begin{table*}[t]
  \centering
  \small
    \caption{Scores on the MTEB v2 (English) benchmark for \adeptmt. $\dagger$ these scores are identical between standard and asymmetric mode.}

  \label{tab:multi-task-results}
  \setlength{\tabcolsep}{4pt}
  \begin{tabular}{l|c|ccccccc|c|c}
    \toprule
                                     & \textbf{Size} & \textbf{Class.} & \textbf{Cluster.} & \textbf{Pair Class.} & \textbf{Rerank}  & \textbf{Retrieval} & \textbf{STS} & \textbf{Summ.} & \textbf{Avg.} & \textbf{Avg. (Datasets)} \\
    \multicolumn{1}{c|}{\# Datasets} &               & 8               & 8                 & 3                    & 2                & 10                 & 9            & 1              &               &                          \\
    \midrule
    \textbf{\adeptmt{} (asym.)}       & 23M           & \ub{76.9}       & \ub{46.5}         & \ub{84.5}            & \ub{\blue{47.3}} & \blue{51.8}        & \ub{82.8}    & \ub{30.9}      & \ub{60.1}     & \ub{64.1}                \\
    \textbf{\adeptmt}                & 23M           & $\dagger$       & $\dagger$         & $\dagger$            & 46.9             & 47.3               & $\dagger$    & $\dagger$      & \ub{59.4}     & \ub{63.0}                \\
    \midrule
    \multicolumn{11}{c}{Comparisons}                                                                                                                                                                                                 \\
    \midrule
    \minilm                          & 23M           & 69.3            & 44.9              & 82.4                 & \ub{47.1}        & 42.9               & 79.0         & 26.0           & 55.9          & 59.0                     \\
    \snowflakexs                     & 23M           & 67.0            & 42.4              & 81.3                 & 45.3             & \ub{52.7}          & 76.2         & 28.0           & 56.1          & 59.8                     \\
    \midrule
    \multicolumn{11}{c}{Teacher}                                                                                                                                                                                                     \\
    \midrule
    \mxbai~                          & 335M          & 79.1            & 47.5              & 87.2                 & 48.1             & 55.4               & 84.4         & 32.6           & 62.0          & 66.3                     \\
    \bottomrule
  \end{tabular}
\end{table*}

Our \adeptmt{} model ranks no.1 on the public leaderboard of this benchmark for models of its size,
at the time of writing. Specifically, our model sets a new state-of-the-art on 5 out of 7 tasks,
as well as on the aggregated average benchmark score, achieving $59.4$. Accordingly, our model
retains 95.8\% of its teacher performance despite the aggressive compression regime.
Tests run on an AWS EC2 \texttt{i3.large} instance show that inference time speed-ups are
of 24.4$\times$ / 23.7$\times$ for docs and queries respectively in this case.
More details about these measurements are in Appendix~\ref{appendix:inference_time_speed}.

When run in asymmetric mode, the performance of our system on reranking and retrieval tasks is
further improved: \adeptmt{} sets a new sota on 6 out of 7 tasks, and its aggregated benchmark score
is increased to $60.1$ (96.9\% of its teacher).

This model also inherits MRL and vector quantization robustness from its teacher.
The performance profiles are similar to those shown in Figure~\ref{fig:mrl_transfer-leaf-ir}, see Appendix~\ref{appendix:mrl_transfer_mt}.


\subsection{Robustness Margins}
\label{sec:systems_robustness}

Throughout this work we consider knowledge distillation as an approximation scheme, where a smaller model attempts
to approximate the embedding map of its teacher. We consequently introduce the $\ell_2$ approximation error in
Eq.~\eqref{eq:training_loss} as the loss to drive our training scheme.

In this subsection we investigate how much approximation error can a retrieval system based on \leaf{} models
sustain without excessive performance degradation.

\begin{figure}[hbtp]
  \includegraphics[width=\linewidth]{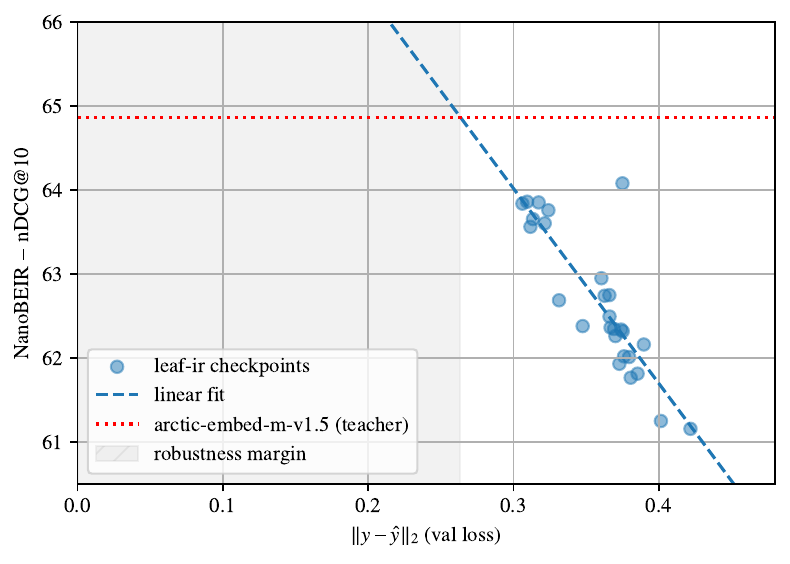}
  \caption{NanoBEIR performance of various \adeptir{} training checkpoints.}
  \label{fig:robustness_margin}
\end{figure}

To this end, Figure~\ref{fig:robustness_margin} shows the downstream performance of several \adeptir{} checkpoints we stored
during training at the end of each epoch\footnote{We removed the checkpoints from the first epoch of each cycle as we found downstream
  performance at high learning rates to display a high degree of variability.}.
In these results, \adeptir{} was used to encode both queries as well as documents.
As can be seen, the lowest average approximation error on the validation set $|I^{\mathrm{val}}|^{-1} \sum_{i \in I^{\mathrm{val}}} \epsilon_i$ we were able to obtain is $\approx 0.3$,
where $\epsilon_i = \|y_i - \yhat_i \|_2$.
Consider that \adeptir's and its teacher embeddings have an $\ell_2$-norm of 1,
meaning that $0 \leq \epsilon_i \leq 2$. As such, the residual approximation error is significant even with our best checkpoints.


On the other hand, Figure~\ref{fig:robustness_margin} also plots a linear trend for our checkpoints.
Besides confirming the intuitive idea that
lower approximation errors lead to scores that are closer to the teacher's reference performance, this curve suggests that
it is not necessary for this approximation error to be 0 for there to be no distinguishable performance difference with
respect to the teacher. We graphically represent the region where we hypothesize this to occur as the shaded region in
Figure~\ref{fig:robustness_margin} and refer to it as the \emph{robustness margin} of the system.
These results are analogous for \adeptmt, see Appendix~\ref{appendix:robustness_adept_mt}.

We believe that the success of our scheme is owed to these robustness margins.

  \section{Conclusion}
\label{sec:conclusions}

This work introduced a lightweight knowledge distillation regime that produces text embedding models \emph{aligned} to their teachers.
These models set a new state-of-the-art on information retrieval and multitask benchmarks respectively, ranking no.1 on the corresponding
leaderboards for models of their size.
Further, to the best of our knowledge ours are the first publicly available models that enable
the flexible asymmetric architecture shown in Figure~\ref{fig:new_paradigm}, and that this architecture can enable further retrieval
performance gains.
We are also the first to show that this distillation procedure automatically
inherits MRL and vector quantization properties from the teacher.



  \section*{Limitations}

  The models we published have only been trained and evaluated on the English language.
  Whether such compact models can accommodate multilinguality (including programming languages) is an open question.

  We also have not investigated the scaling of our procedure, i.e., whether larger models can be synthesized as effectively.
  Many recent embedding models use decoders instead of encoders, and whether the technique needs to be adapted in this
  case also needs further investigation.

  Finally, we have adapted and incorporated several other distillation losses in our experiments, such as MiniLM, TinyBERT and
  DistilBERT, as reported in Appendix~\ref{appendix:kd_ablation}, but weren't able to
  materialize concrete improvements in effectiveness.
  This is the case despite the fact that these additions take advantage
  of more information from the teacher, in the form of its internal representations and attention matrices.
  More experimentation may shed further light on this counterintuitive result.


  \section*{Acknowledgments}
  We are grateful to Henry Weller, Prakul Agarwal, Seny Kamara and John Partridge for
  the several illuminating discussions; they directly contributed to ensuring the practical
  relevance of this work. We also thank James Gentile for reviewing the paper and providing the initial context
  that prompted this research work, as well as Hong Liu for helping with references and
  reviewing the paper.

  \bibliography{references}

@article{vaswani2017attention,
  title   = {Attention is all you need},
  author  = {Vaswani, Ashish and Shazeer, Noam and Parmar, Niki and Uszkoreit, Jakob and Jones, Llion and Gomez, Aidan N and Kaiser, {\L}ukasz and Polosukhin, Illia},
  journal = {Advances in neural information processing systems},
  volume  = {30},
  year    = {2017}
}

@article{minilm,
  title   = {Minilm: Deep self-attention distillation for task-agnostic compression of pre-trained transformers},
  author  = {Wang, Wenhui and Wei, Furu and Dong, Li and Bao, Hangbo and Yang, Nan and Zhou, Ming},
  journal = {Advances in neural information processing systems},
  volume  = {33},
  pages   = {5776--5788},
  year    = {2020}
}

@article{beir,
  title   = {Beir: A heterogenous benchmark for zero-shot evaluation of information retrieval models},
  author  = {Thakur, Nandan and Reimers, Nils and R{\"u}ckl{\'e}, Andreas and Srivastava, Abhishek and Gurevych, Iryna},
  journal = {arXiv preprint arXiv:2104.08663},
  year    = {2021}
}

@article{arctic_v1,
  title   = {Arctic-embed: Scalable, efficient, and accurate text embedding models},
  author  = {Merrick, Luke and Xu, Danmei and Nuti, Gaurav and Campos, Daniel},
  journal = {arXiv preprint arXiv:2405.05374},
  year    = {2024}
}

@misc{bm25_benchmark,
  author       = {Xhluca},
  title        = {BM25 Benchmarks},
  year         = {2023},
  howpublished = {\url{https://github.com/xhluca/bm25-benchmarks/blob/22df0ccccc083d2985a5b7e55d3ecd1764d4f9ed/README.md?plain=1\#L220}},
  note         = {Accessed: 2025-05-07}
}

@misc{colbert_v2,
  title         = {ColBERTv2: Effective and Efficient Retrieval via Lightweight Late Interaction},
  author        = {Keshav Santhanam and Omar Khattab and Jon Saad-Falcon and Christopher Potts and Matei Zaharia},
  year          = {2022},
  eprint        = {2112.01488},
  archiveprefix = {arXiv},
  primaryclass  = {cs.IR},
  url           = {https://arxiv.org/abs/2112.01488}
}

@article{mteb_eng_v2,
  author    = {Kenneth Enevoldsen and Isaac Chung and Imene Kerboua and Márton Kardos and Ashwin Mathur and David Stap and Jay Gala and Wissam Siblini and Dominik Krzemiński and Genta Indra Winata and Saba Sturua and Saiteja Utpala and Mathieu Ciancone and Marion Schaeffer and Gabriel Sequeira and Diganta Misra and Shreeya Dhakal and Jonathan Rystrøm and Roman Solomatin and Ömer Çağatan and Akash Kundu and Martin Bernstorff and Shitao Xiao and Akshita Sukhlecha and Bhavish Pahwa and Rafał Poświata and Kranthi Kiran GV and Shawon Ashraf and Daniel Auras and Björn Plüster and Jan Philipp Harries and Loïc Magne and Isabelle Mohr and Mariya Hendriksen and Dawei Zhu and Hippolyte Gisserot-Boukhlef and Tom Aarsen and Jan Kostkan and Konrad Wojtasik and Taemin Lee and Marek Šuppa and Crystina Zhang and Roberta Rocca and Mohammed Hamdy and Andrianos Michail and John Yang and Manuel Faysse and Aleksei Vatolin and Nandan Thakur and Manan Dey and Dipam Vasani and Pranjal Chitale and Simone Tedeschi and Nguyen Tai and Artem Snegirev and Michael Günther and Mengzhou Xia and Weijia Shi and Xing Han Lù and Jordan Clive and Gayatri Krishnakumar and Anna Maksimova and Silvan Wehrli and Maria Tikhonova and Henil Panchal and Aleksandr Abramov and Malte Ostendorff and Zheng Liu and Simon Clematide and Lester James Miranda and Alena Fenogenova and Guangyu Song and Ruqiya Bin Safi and Wen-Ding Li and Alessia Borghini and Federico Cassano and Hongjin Su and Jimmy Lin and Howard Yen and Lasse Hansen and Sara Hooker and Chenghao Xiao and Vaibhav Adlakha and Orion Weller and Siva Reddy and Niklas Muennighoff},
  doi       = {10.48550/arXiv.2502.13595},
  journal   = {arXiv preprint arXiv:2502.13595},
  publisher = {arXiv},
  title     = {MMTEB: Massive Multilingual Text Embedding Benchmark},
  url       = {https://arxiv.org/abs/2502.13595},
  year      = {2025}
}

@inproceedings{fineweb,
  title     = {The FineWeb Datasets: Decanting the Web for the Finest Text Data at Scale},
  author    = {Guilherme Penedo and Hynek Kydl{\'\i}{\v{c}}ek and Loubna Ben allal and Anton Lozhkov and Margaret Mitchell and Colin Raffel and Leandro Von Werra and Thomas Wolf},
  booktitle = {The Thirty-eight Conference on Neural Information Processing Systems Datasets and Benchmarks Track},
  year      = {2024},
  url       = {https://openreview.net/forum?id=n6SCkn2QaG}
}

@misc{500k_words,
  title     = {Dataset containing 479k English words},
  url       = {https://www.kaggle.com/ds/4075094},
  doi       = {10.34740/KAGGLE/DS/4075094},
  author    = {Kaggle},
  publisher = {Kaggle},
  year      = {2023}
}

@article{msmarco,
  author        = {Tri Nguyen and
                   Mir Rosenberg and
                   Xia Song and
                   Jianfeng Gao and
                   Saurabh Tiwary and
                   Rangan Majumder and
                   Li Deng},
  title         = {{MS} {MARCO:} {A} Human Generated MAchine Reading COmprehension Dataset},
  journal       = {CoRR},
  volume        = {abs/1611.09268},
  year          = {2016},
  url           = {http://arxiv.org/abs/1611.09268},
  archiveprefix = {arXiv},
  eprint        = {1611.09268},
  bibsource     = {dblp computer science bibliography, https://dblp.org}
}

@article{amazonqa,
  title   = {Amazonqa: A review-based question answering task},
  author  = {Gupta, Mansi and Kulkarni, Nitish and Chanda, Raghuveer and Rayasam, Anirudha and Lipton, Zachary C},
  journal = {arXiv preprint arXiv:1908.04364},
  year    = {2019}
}

@inproceedings{pubmedqa,
  title     = {PubMedQA: A Dataset for Biomedical Research Question Answering},
  author    = {Jin, Qiao and Dhingra, Bhuwan and Liu, Zhengping and Cohen, William and Lu, Xinghua},
  booktitle = {Proceedings of the 2019 Conference on Empirical Methods in Natural Language Processing and the 9th International Joint Conference on Natural Language Processing (EMNLP-IJCNLP)},
  pages     = {2567--2577},
  year      = {2019}
}

@article{triviaqa,
  author        = {{Joshi}, Mandar and {Choi}, Eunsol and {Weld},
                   Daniel and {Zettlemoyer}, Luke},
  title         = {{triviaqa: A Large Scale Distantly Supervised Challenge Dataset for Reading Comprehension}},
  journal       = {arXiv e-prints},
  year          = 2017,
  eid           = {arXiv:1705.03551},
  pages         = {arXiv:1705.03551},
  archiveprefix = {arXiv},
  eprint        = {1705.03551}
}

@inproceedings{bert_devlin,
  title     = {Bert: Pre-training of deep bidirectional transformers for language understanding},
  author    = {Devlin, Jacob and Chang, Ming-Wei and Lee, Kenton and Toutanova, Kristina},
  booktitle = {Proceedings of the 2019 conference of the North American chapter of the association for computational linguistics: human language technologies, volume 1 (long and short papers)},
  pages     = {4171--4186},
  year      = {2019}
}

@inproceedings{tinybert,
  title     = {{T}iny{BERT}: Distilling {BERT} for Natural Language Understanding},
  author    = {Jiao, Xiaoqi  and
               Yin, Yichun  and
               Shang, Lifeng  and
               Jiang, Xin  and
               Chen, Xiao  and
               Li, Linlin  and
               Wang, Fang  and
               Liu, Qun},
  editor    = {Cohn, Trevor  and
               He, Yulan  and
               Liu, Yang},
  booktitle = {Findings of the Association for Computational Linguistics: EMNLP 2020},
  month     = nov,
  year      = {2020},
  address   = {Online},
  publisher = {Association for Computational Linguistics},
  url       = {https://aclanthology.org/2020.findings-emnlp.372/},
  doi       = {10.18653/v1/2020.findings-emnlp.372},
  pages     = {4163--4174}
}

@misc{distilbert,
  title         = {DistilBERT, a distilled version of BERT: smaller, faster, cheaper and lighter},
  author        = {Victor Sanh and Lysandre Debut and Julien Chaumond and Thomas Wolf},
  year          = {2020},
  eprint        = {1910.01108},
  archiveprefix = {arXiv},
  primaryclass  = {cs.CL},
  url           = {https://arxiv.org/abs/1910.01108}
}

@misc{distilbert_implementation,
  author = {Hugging Face},
  title  = {DistilBERT},
  year   = {2023},
  url    = {https://github.com/huggingface/transformers-research-projects/blob/362a490dc36e91359fe76a7a707dc29e663196b2/distillation/distiller.py#L434},
  note   = {Accessed: 2025-05-22}
}

@misc{splade_v2_distillation,
  title         = {Improving Efficient Neural Ranking Models with Cross-Architecture Knowledge Distillation},
  author        = {Sebastian Hofst{\"a}tter and Sophia Althammer and Michael Schr{\"o}der and Mete Sertkan and Allan Hanbury},
  year          = {2020},
  eprint        = {2010.02666},
  archiveprefix = {arXiv},
  primaryclass  = {cs.IR}
}

@inbook{splade_v1,
  author    = {Formal, Thibault and Piwowarski, Benjamin and Clinchant, St\'{e}phane},
  title     = {SPLADE: Sparse Lexical and Expansion Model for First Stage Ranking},
  year      = {2021},
  isbn      = {9781450380379},
  publisher = {Association for Computing Machinery},
  address   = {New York, NY, USA},
  url       = {https://doi.org/10.1145/3404835.3463098},
  booktitle = {Proceedings of the 44th International ACM SIGIR Conference on Research and Development in Information Retrieval},
  pages     = {2288–2292},
  numpages  = {5}
}

@misc{openai_embed,
  title         = {Text and Code Embeddings by Contrastive Pre-Training},
  author        = {Arvind Neelakantan and Tao Xu and Raul Puri and Alec Radford and Jesse Michael Han and Jerry Tworek and Qiming Yuan and Nikolas Tezak and Jong Wook Kim and Chris Hallacy and Johannes Heidecke and Pranav Shyam and Boris Power and Tyna Eloundou Nekoul and Girish Sastry and Gretchen Krueger and David Schnurr and Felipe Petroski Such and Kenny Hsu and Madeleine Thompson and Tabarak Khan and Toki Sherbakov and Joanne Jang and Peter Welinder and Lilian Weng},
  year          = {2022},
  eprint        = {2201.10005},
  archiveprefix = {arXiv},
  primaryclass  = {cs.CL},
  url           = {https://arxiv.org/abs/2201.10005}
}

@misc{arctic_embed_v2,
  title         = {Arctic-Embed 2.0: Multilingual Retrieval Without Compromise},
  author        = {Puxuan Yu and Luke Merrick and Gaurav Nuti and Daniel Campos},
  year          = {2024},
  eprint        = {2412.04506},
  archiveprefix = {arXiv},
  primaryclass  = {cs.CL},
  url           = {https://arxiv.org/abs/2412.04506}
}

@inproceedings{grit,
  title     = {Generative Representational Instruction Tuning},
  author    = {Niklas Muennighoff and Hongjin SU and Liang Wang and Nan Yang and Furu Wei and Tao Yu and Amanpreet Singh and Douwe Kiela},
  booktitle = {The Thirteenth International Conference on Learning Representations},
  year      = {2025},
  url       = {https://openreview.net/forum?id=BC4lIvfSzv}
}

@inproceedings{contrastive_batch_sizes_simclr,
  author    = {Chen, Ting and Kornblith, Simon and Norouzi, Mohammad and Hinton, Geoffrey},
  title     = {A simple framework for contrastive learning of visual representations},
  year      = {2020},
  publisher = {JMLR.org},
  booktitle = {Proceedings of the 37th International Conference on Machine Learning},
  articleno = {149},
  numpages  = {11},
  series    = {ICML'20}
}

@inproceedings{mrl,
  author    = {Kusupati, Aditya and Bhatt, Gantavya and Rege, Aniket and Wallingford, Matthew and Sinha, Aditya and Ramanujan, Vivek and Howard-Snyder, William and Chen, Kaifeng and Kakade, Sham and Jain, Prateek and Farhadi, Ali},
  booktitle = {Advances in Neural Information Processing Systems},
  editor    = {S. Koyejo and S. Mohamed and A. Agarwal and D. Belgrave and K. Cho and A. Oh},
  pages     = {30233--30249},
  publisher = {Curran Associates, Inc.},
  title     = {Matryoshka Representation Learning},
  url       = {https://proceedings.neurips.cc/paper_files/paper/2022/file/c32319f4868da7613d78af9993100e42-Paper-Conference.pdf},
  volume    = {35},
  year      = {2022}
}

@book{nielsen_throughput,
  title     = {Usability engineering},
  author    = {Nielsen, Jakob},
  year      = {1994},
  publisher = {Morgan Kaufmann}
}

@article{mobilebert,
  title   = {Mobilebert: a compact task-agnostic bert for resource-limited devices},
  author  = {Sun, Zhiqing and Yu, Hongkun and Song, Xiaodan and Liu, Renjie and Yang, Yiming and Zhou, Denny},
  journal = {arXiv preprint arXiv:2004.02984},
  year    = {2020}
}

@inproceedings{reimers_making_monolingual_models_multilingual,
  title     = {Making Monolingual Sentence Embeddings Multilingual using Knowledge Distillation},
  author    = {Reimers, Nils  and
               Gurevych, Iryna},
  editor    = {Webber, Bonnie  and
               Cohn, Trevor  and
               He, Yulan  and
               Liu, Yang},
  booktitle = {Proceedings of the 2020 Conference on Empirical Methods in Natural Language Processing (EMNLP)},
  month     = nov,
  year      = {2020},
  address   = {Online},
  publisher = {Association for Computational Linguistics},
  url       = {https://aclanthology.org/2020.emnlp-main.365/},
  doi       = {10.18653/v1/2020.emnlp-main.365},
  pages     = {4512--4525}
}

@article{kale_model,
  title   = {Quick dense retrievers consume kale: Post training kullback leibler alignment of embeddings for asymmetrical dual encoders},
  author  = {Campos, Daniel and Magnani, Alessandro and Zhai, ChengXiang},
  journal = {arXiv preprint arXiv:2304.01016},
  year    = {2023}
}

@inproceedings{instructor_instruction_prompts_embedding_models,
  title     = {One Embedder, Any Task: Instruction-Finetuned Text Embeddings},
  author    = {Su, Hongjin  and
               Shi, Weijia  and
               Kasai, Jungo  and
               Wang, Yizhong  and
               Hu, Yushi  and
               Ostendorf, Mari  and
               Yih, Wen-tau  and
               Smith, Noah A.  and
               Zettlemoyer, Luke  and
               Yu, Tao},
  editor    = {Rogers, Anna  and
               Boyd-Graber, Jordan  and
               Okazaki, Naoaki},
  booktitle = {Findings of the Association for Computational Linguistics: ACL 2023},
  month     = jul,
  year      = {2023},
  address   = {Toronto, Canada},
  publisher = {Association for Computational Linguistics},
  url       = {https://aclanthology.org/2023.findings-acl.71/},
  doi       = {10.18653/v1/2023.findings-acl.71},
  pages     = {1102--1121}
}

@inproceedings{tinybert_simplified,
  author    = {Chen, Xuanang and He, Ben and Hui, Kai and Sun, Le and Sun, Yingfei},
  title     = {Simplified TinyBERT: Knowledge Distillation for Document Retrieval},
  year      = {2021},
  isbn      = {978-3-030-72239-5},
  publisher = {Springer-Verlag},
  address   = {Berlin, Heidelberg},
  url       = {https://doi.org/10.1007/978-3-030-72240-1_21},
  doi       = {10.1007/978-3-030-72240-1_21},
  booktitle = {Advances in  Information Retrieval: 43rd European Conference on IR Research, ECIR 2021, Virtual Event, March 28 – April 1, 2021, Proceedings, Part II},
  pages     = {241–248},
  numpages  = {8}
}

@inproceedings{tite_paper,
  author    = {Schlatt, Ferdinand and Hagen, Tim and Potthast, Martin and Hagen, Matthias},
  title     = {TITE: Token-Independent Text Encoder for Information Retrieval},
  year      = {2025},
  isbn      = {9798400715921},
  publisher = {Association for Computing Machinery},
  address   = {New York, NY, USA},
  url       = {https://doi.org/10.1145/3726302.3730094},
  doi       = {10.1145/3726302.3730094},
  booktitle = {Proceedings of the 48th International ACM SIGIR Conference on Research and Development in Information Retrieval},
  pages     = {2493–2503},
  numpages  = {11},
  location  = {Padua, Italy},
  series    = {SIGIR '25}
}

@inproceedings{embedding-converter,
    title = "Embedding-Converter: A Unified Framework for Cross-Model Embedding Transformation",
    author = "Yoon, Jinsung  and
      Arik, Sercan O",
    editor = "Che, Wanxiang  and
      Nabende, Joyce  and
      Shutova, Ekaterina  and
      Pilehvar, Mohammad Taher",
    booktitle = "Proceedings of the 63rd Annual Meeting of the Association for Computational Linguistics (Volume 1: Long Papers)",
    month = jul,
    year = "2025",
    address = "Vienna, Austria",
    publisher = "Association for Computational Linguistics",
    url = "https://aclanthology.org/2025.acl-long.1237/",
    doi = "10.18653/v1/2025.acl-long.1237",
    pages = "25464--25482",
    ISBN = "979-8-89176-251-0"
}

@misc{sentencetransformers_distillation,
  author = {{Hugging Face}},
  title = {Model Distillation Example for Sentence Transformers},
  year = {2024},
  howpublished = {\url{https://github.com/huggingface/sentence-transformers/blob/main/examples/sentence_transformer/training/distillation/model_distillation.py}},
  note = {Accessed: 2025-12-18}
}

@inproceedings{emo_distillation,
    title = "{EMO}: Embedding Model Distillation via Intra-Model Relation and Optimal Transport Alignments",
    author = "Truong, Minh-Phuc  and
      Vu, Hai An  and
      Vu, Tu  and
      Diep, Nguyen Thi Ngoc  and
      Van, Linh Ngo  and
      Nguyen, Thien Huu  and
      Le, Trung",
    editor = "Christodoulopoulos, Christos  and
      Chakraborty, Tanmoy  and
      Rose, Carolyn  and
      Peng, Violet",
    booktitle = "Proceedings of the 2025 Conference on Empirical Methods in Natural Language Processing",
    month = nov,
    year = "2025",
    address = "Suzhou, China",
    publisher = "Association for Computational Linguistics",
    url = "https://aclanthology.org/2025.emnlp-main.385/",
    doi = "10.18653/v1/2025.emnlp-main.385",
    pages = "7605--7617",
    ISBN = "979-8-89176-332-6"
}

@inproceedings{m3_paper,
    title = "{M}3-Embedding: Multi-Linguality, Multi-Functionality, Multi-Granularity Text Embeddings Through Self-Knowledge Distillation",
    author = "Chen, Jianlyu  and
      Xiao, Shitao  and
      Zhang, Peitian  and
      Luo, Kun  and
      Lian, Defu  and
      Liu, Zheng",
    editor = "Ku, Lun-Wei  and
      Martins, Andre  and
      Srikumar, Vivek",
    booktitle = "Findings of the Association for Computational Linguistics: ACL 2024",
    month = aug,
    year = "2024",
    address = "Bangkok, Thailand",
    publisher = "Association for Computational Linguistics",
    url = "https://aclanthology.org/2024.findings-acl.137/",
    doi = "10.18653/v1/2024.findings-acl.137",
    pages = "2318--2335"
}

@misc{qwen3_vl_embedding_and_reranker,
      title={Qwen3-VL-Embedding and Qwen3-VL-Reranker: A Unified Framework for State-of-the-Art Multimodal Retrieval and Ranking},
      author={Mingxin Li and Yanzhao Zhang and Dingkun Long and Keqin Chen and Sibo Song and Shuai Bai and Zhibo Yang and Pengjun Xie and An Yang and Dayiheng Liu and Jingren Zhou and Junyang Lin},
      year={2026},
      eprint={2601.04720},
      archivePrefix={arXiv},
      primaryClass={cs.CL},
      url={https://arxiv.org/abs/2601.04720},
}

  \appendix

\section{Training Setup}

\subsection{Pooling Layer}
\label{appendix:pooling_layer_ablation}

\begin{figure}[hbtp]
  \centering
  \includegraphics[width=0.48\textwidth]{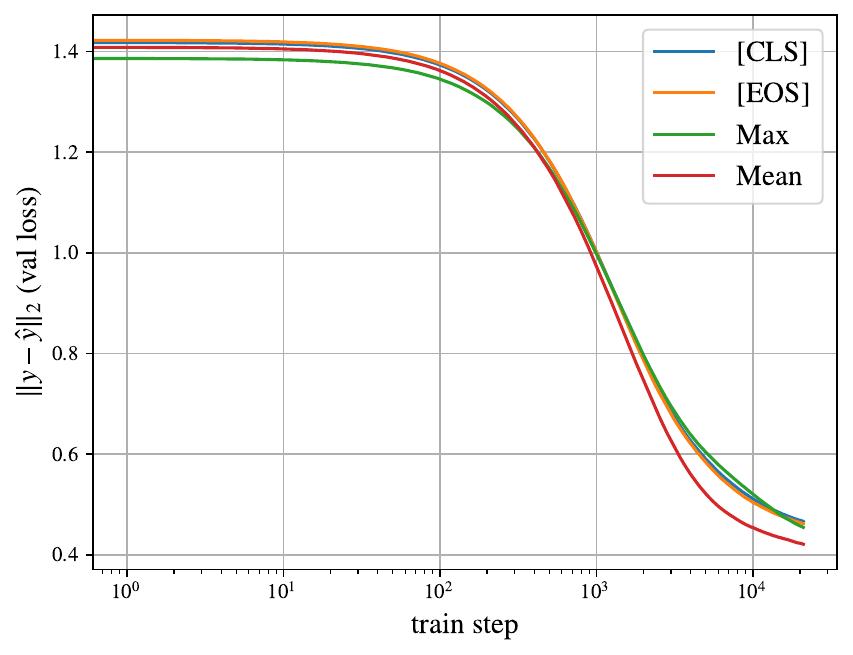}
  \caption{Pooling layer comparison. Mean pooling achieves the lowest (best) final validation loss.}
  \label{fig:pooling_layer_comparison}
\end{figure}

In Figure~\ref{fig:pooling_layer_comparison} a comparison of different popular pooling layers is shown.
These are obtained by running one training epoch of \adeptir{} on the entire training dataset.
We observe that mean pooling produces lower
(better) validation scores than \texttt{[CLS]}, \texttt{[EOS]} and max pooling at the end of the training epoch.



\subsection{Learning Rate Decay Schedule}
\label{appendix:learn_rate_decay_schedule}

\begin{figure}[hbtp]
  \centering
  \caption{Comparison of learning rate decay schedules. Linear decay ends with the lowest (best) final validation loss when the training data is
    most abundant in this comparison, more closely resembling our production runs.}
  \label{fig:learn_rate_decay_schedule_comparison}
  \includegraphics[width=1\linewidth]{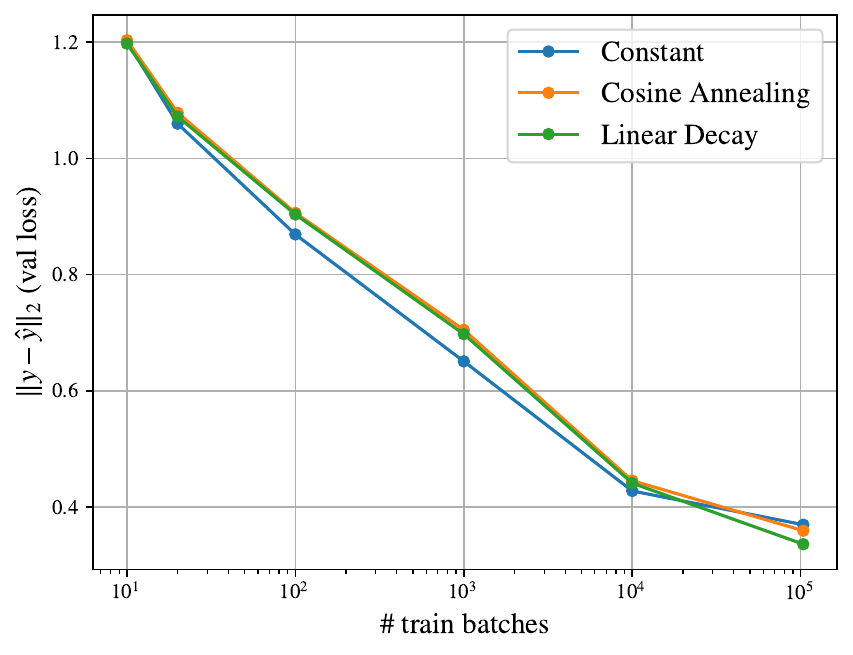}
\end{figure}

Figure~\ref{fig:learn_rate_decay_schedule_comparison} shows a comparison of different popular learning rate decay schedules.
Each point reflects a run of 10 training epochs of \adeptir, where each epoch contains a number of training batches reported on the x-axis. On the y-axis
we have the final validation loss. For the constant variation, we keep lr=$1e-4$ throughout the 10 epochs. For linear decay, we decrease the learning
rate at the end of each epoch, from lr=$1e-4$ to lr=$1e-5$ at the 10th epoch. Cosine annealing steps the learning rate down each training
step.

Constant learning rate achieves the lowest (best) validation scores at low training data regimes, while linear decay has the best loss
amongst the rate schedules we compared in the presence of more training data, when each epoch consist of 100k training batches, more
closely matching our production runs.

\subsection{Batch Size Selection}
\label{appendix:batch_size_selection}

Table~\ref{tab:batch_size_selection} reports our results when running one training epoch of \adeptir{} under different batch sizes.
The table contains both batch size as well as total number of batches: the amount of training data is kept constant so smaller
batch sizes lead to a larger number of training batches. The wall clock time to complete the epoch is also reported; training
is conducted on a single A100 40GB GPU. The validation loss reported is the loss we measure at the end of the training epoch.

\begin{table}[ht]
  \centering
  \small
  \caption{Training metrics for different batch sizes. Batch size of 32 has the best trade-off between epoch completion time and dev loss at the end of the epoch.}
  \label{tab:batch_size_selection}
  \begin{tabular}{cccc}
    \toprule
    \textbf{Batch Size} & \textbf{Time} & \textbf{Val Loss $\ell_2$} & \textbf{\# Batches} \\
    \midrule
    16                  & 4h 19min      & 0.4194                     & 400k                \\
    32                  & 3h 05min      & 0.4214                     & 200k                \\
    64                  & 2h 51min      & 0.4301                     & 100k                \\
    128                 & 2h 40min      & 0.4390                     & 50k                 \\
    256                 & 2h 32min      & 0.4593                     & 25k                 \\
    \bottomrule
  \end{tabular}
\end{table}

One can observe that the final losses for batch sizes 16 and 32 are similar, but as we further increase the batch sizes
the final loss degrades more and more, while the training wall clock time does not improve as much.
We find the sweetspot between wall clock time and final validation loss at a batch size of 32.

This result shows that our scheme generally favors more steps with smaller batch sizes, rather than the other way around.
This makes intuitive sense as the model needs to be completely re-oriented to match the embedding vector space of the target
teacher model.


\section{Alternative Knowledge Distillation Approaches}
\label{appendix:kd_ablation}


We implemented the losses of three popular knowledge distillation frameworks:
MiniLM \cite{minilm}, TinyBERT \cite{tinybert} and DistilBERT \cite{distilbert}.

All of these frameworks are designed for the distillation of language models, i.e., transformer architectures
equipped with a language modelling head. Text embedding models based on transformers follow the standard architecture
discussed in Section~\ref{sec:approach} and usually remove the language modelling head and replace it
with a pooling layer and an optional normalization layer. We additionally add a linear layer
$\wout \in \R^{\dimadept \times \dimembed} $, to map \adept's embedding dimension $\dimadept$ to the desired
teacher output dimension $\dimembed$. As shown below, none of the losses proposed by these knowledge distillation
procedures would entail a training signal for $\wout$. To address this, we combined each loss with our
$\ell_2$ distillation loss in Eq.~\eqref{eq:training_loss}. The inclusion of this loss component is also
necessary to ensure that the models distilled are \emph{aligned} to their teacher, a core goal of this work.

Further, as proposed in their original papers, these knowledge distillation schemes require the
tokenizers of the teacher and the student to match. MiniLM and TinyBERT further require the number of attention
heads to match. Our scheme does not have these restrictions.

\begin{figure*}[htbp]
  \centering
  \begin{subfigure}[b]{0.48\textwidth}
    \centering
    \includegraphics[width=\textwidth]{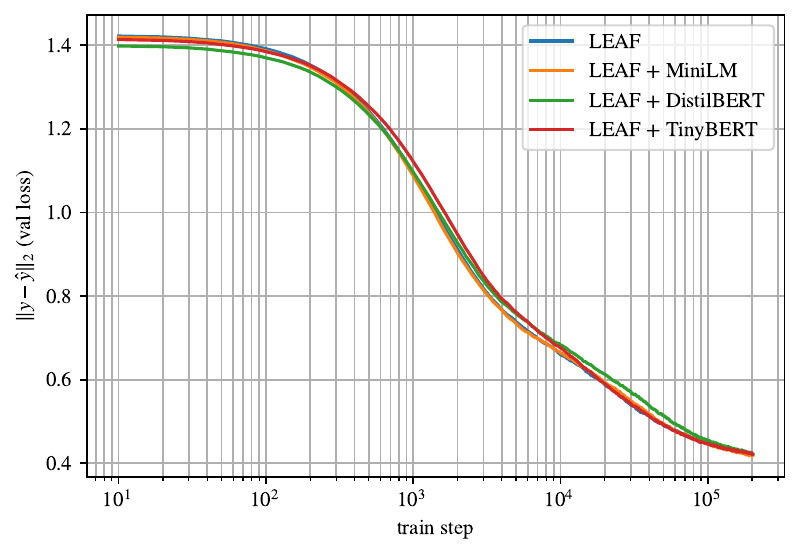}
    \caption{Validation loss over the entire 1st epoch}
    \label{fig:kd_ablation}
  \end{subfigure}
  \hfill
  \begin{subfigure}[b]{0.48\textwidth}
    \centering
    \includegraphics[width=\textwidth]{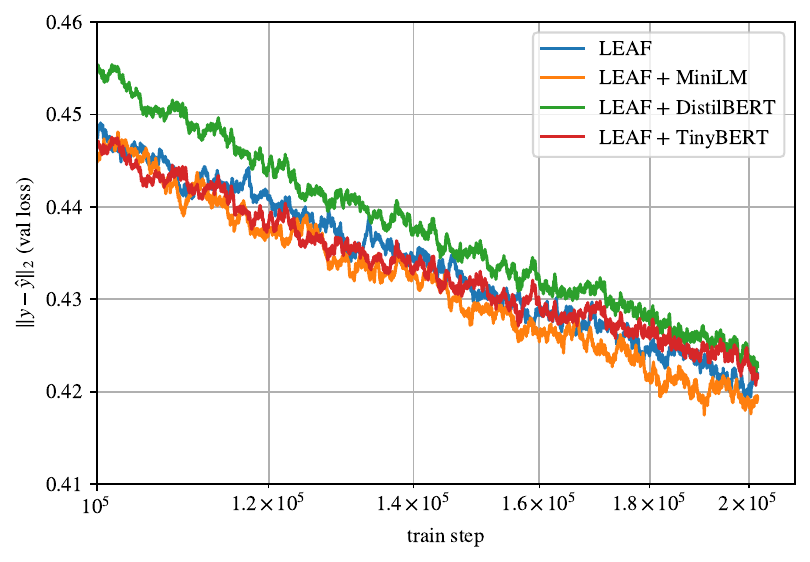}
    \caption{Zoom in on the last 100k steps}
    \label{fig:kd_ablation_zoom_zoom}
  \end{subfigure}
  \caption{Comparison of the validation losses observed over the first training epoch under different knowledge distillation
    approaches discussed in Appendix~\ref{appendix:kd_ablation}. The difference in validation losses measured at the end of the
    first training epoch is insubstantial.}
  \label{fig:kd_ablation_combined}
\end{figure*}

We run one epoch with each one of these distillation schemes when combined with our
output alignment loss Eq.~\eqref{eq:training_loss}. Figure~\ref{fig:kd_ablation} shows the validation loss curves
over this epoch, while Figure~\ref{fig:kd_ablation_zoom_zoom} zooms in on the last 100k steps. The final
approximation errors $\|y - \yhat \|_2$ on the validation set we obtain with all these schemes is very similar, as
shown in the Figures.


The downstream performance on NanoBEIR reported in Table~\ref{tab:kd_nanomsmarco}, however, suggests
that none of these additions produces better results than working with just the loss in Eq.~\eqref{eq:training_loss}.

\begin{table}[ht]
  \centering
  \caption{Comparison of different knowledge distillation losses measured on the NanoMSMARCO dataset, after one training epoch.
    These results indicate that additional training signals leveraging Transformer's internal representations may not lead to better downstream results.}
  \label{tab:kd_nanomsmarco}
  \begin{tabular}{lc}
    \toprule
    \textbf{Loss}                                                      & \textbf{nDCG@10} \\
    \midrule
    $\mathcal{L}_{\mathrm{\LEAF}}$                                     & 60.7             \\
    $\mathcal{L}_{\mathrm{\LEAF}} + \mathcal{L}_{\mathrm{MiniLM}}$     & 54.9             \\
    $\mathcal{L}_{\mathrm{\LEAF}} + \mathcal{L}_{\mathrm{TinyBERT}}$   & 53.7             \\
    $\mathcal{L}_{\mathrm{\LEAF}} + \mathcal{L}_{\mathrm{DistilBERT}}$ & 55.3             \\
    \bottomrule
  \end{tabular}
\end{table}

Below we provide more details about how we adapted these methods in our implementation.


\subsection{MiniLM}
\label{sec:minilm_ablation}
We note that for this training procedure to work as proposed in \cite{minilm}, the number of attention heads
in the student and the teacher models have to match. This is not required by our loss~\eqref{eq:training_loss}.

Given the value vector $\mathrm{V}_{l,a} \in \R^{T \times C}$ for layer $l$ and attention head $a$, where $T \in \N$ is the
sequence length and $C \in \N$ is the number of channels in the value vector of attention $a$; typically, $C = d/A$, where
$d \in \N$ is the hidden dimension of the Transformer and $A \in \N$ is the number of heads. We disregard the batch dimension in this analysis.
The \emph{value relation matrix} $\mathrm{VR}_{l,a} \in \R^{T \times T}$ is then defined as
\begin{equation}
  \label{eq:value_relation}
  \mathrm{VR}_{l,a} = \mathrm{softmax} \left( \frac{\mathrm{V}_{l,a} \cdot \mathrm{V}_{l,a}\tr}{\sqrt{C}} \right).
\end{equation}
Let $L$ and $L'$ be the number of transformer layers in the teacher and student models respectively. Then,
\begin{equation}
  \label{eq:minilm_loss_vr}
  \mathcal{L}_{\mathrm{VR}} = \frac{1}{A T} \sum_{t=1}^T \sum_{a = 1}^{A} \mathcal{D}_{\mathrm{KL}}(\mathrm{VR}^\tau_{L,a} \left|\right| \mathrm{VR}^\sigma_{L',a})_t,
\end{equation}
where $\mathcal{D}_{\mathrm{KL}}$ is the KL-divergence between the value relation~\eqref{eq:value_relation} of the teacher $\tau$ and
student $\sigma$.
According to Eq.~\eqref{eq:minilm_loss_vr}, loss is computed only for the last transformer layers of the respective networks $\Ll$ and $\Lp$, and averaged over the attention
heads and input tokens.

An additional loss term is added to align the attention patterns of the two transformer backbones. Namely, let
\begin{equation}
  \label{eq:transformer_attention}
  \mathrm{Att}_{l,a} = \mathrm{softmax} \left( \frac{\mathrm{K}_{l,a} \cdot \mathrm{Q}_{l,a}\tr}{\sqrt{C}} \right)
\end{equation}
where $\mathrm{K}_{l,a} \in \R^{T \times C}$ are the keys and $\mathrm{Q}_{l,a} \in \R^{T \times C}$ are the
queries in head $a$ of layer $l$ of the transformer. $\mathrm{Att}_{l,a} \in \R^{T \times T}$ is thus the attention
matrix in the given transformer head. Then, similarly to Eq.\eqref{eq:minilm_loss_vr}, we define
\begin{equation}
  \label{eq:minilm_loss_a}
  \mathcal{L}_{\mathrm{Att}} = \frac{1}{A T} \sum_{t=1}^T \sum_{a = 1}^{A} \mathcal{D}_{\mathrm{KL}}(\mathrm{Att}^\tau_{L,a} \left|\right| \mathrm{Att}^\sigma_{L',a})_t,
\end{equation}
as the KL-divergence between the student's $\sigma$ and the teacher $\tau$ attentions in the last transformer layer,
averaged over the attention heads and input tokens.

The loss we adopt to produce the results in Table~\ref{tab:kd_nanomsmarco} is then
\begin{equation}
  \mathcal{L}_{\mathrm{MiniLM}} =  \mathcal{L}_{\mathrm{VR}} + \mathcal{L}_{\mathrm{Att}}.
\end{equation}

\subsection{TinyBERT}

TinyBERT's \cite{tinybert} proposed knowledge distillation scheme entails a soft cross-entropy loss between teacher and student output logits.
However, as remarked earlier, we discard the language modelling head in our architecture, and hence discard this loss component too.

We next discuss our implementation of the other loss components proposed in TinyBERT.

Let $h_{l,t} \in \R^\dimembed$ be the hidden state at the $l$-th layer and at the $t$-th token.
As a convention, we denote $h_{0,t}$ as the hidden state before the first transformer layer; in BERT, $h_{0,t}$ is the
superposition of the input and positional embeddings.

To align hidden states, TinyBERT proposes the following loss:
\begin{equation}
  \mathcal{L}_{\mathrm{Hidn}} = \frac{1}{T L} \sum_{t=1}^T \sum_{l=1}^{L'}  \mathrm{MSE}( h^\tau_{g(l),t} - W^{\mathrm{map}} h^\sigma_{l,t} ),
\end{equation}
where $g: \N \rightarrow \N$ maps a hidden layer index in the student model to a corresponding hidden layer in the teacher and
is such that $g(0) = 0$ and $g(\Lp) = \Ll$. As proposed in \cite{minilm}, we use $g(l) = \lfloor\frac{\Ll}{\Lp}\rfloor l$.
$W^{\mathrm{map}} \in \R^{\dimadept \times \dimembed}$ maps the student's internal representations, which are often of smaller dimensions, into the teacher ones.

Similar to Section~\ref{sec:minilm_ablation}, TinyBERT also adopts a loss to align the attention patterns as follows:
\begin{equation}
  \mathcal{L}_{\mathrm{Att}} = \frac{1}{A L} \sum_{a = 1}^{A} \sum_{l=1}^{L'} \mathrm{MSE}( \widetilde{\mathrm{Att}}^\tau_{g(l),a}, \widetilde{\mathrm{Att}}^\sigma_{l,a} ),
\end{equation}
where $\widetilde{\mathrm{Att}}$ is the attention matrix defined in Eq.~\eqref{eq:transformer_attention} but without $\mathrm{softmax(\cdot)}$.
Because of this loss component, TinyBERT \cite{tinybert} requires the number of
heads in the student's and teacher's transformer layers to match, like with the MiniLM loss discussed in Section~\ref{sec:minilm_ablation}.

The aggregated loss we use to implement this knowledge distillation strategy is then
\begin{equation}
  \mathcal{L}_\mathrm{TinyBERT} = \mathcal{L}_{\mathrm{Hidn}} + \mathcal{L}_{\mathrm{Att}}.
\end{equation}


\subsection{DistilBERT}

DistilBERT \cite{distilbert} employs a combined loss on output logits and cosine similarity on internal
states. As before, we discard the loss on the logits because we do not have a language modelling head.
The paper does not specify how to compute the loss component related to the hidden states when the
number of layers do not match. The implementation at \cite{distilbert_implementation} suggests that this
loss is only applied to the last layer. It also requires hidden state dimensions to match. We compensate
for this latter limitation by applying a linear map like in Eq.~\eqref{eq:minilm_loss_a}.

We thus have
\begin{equation}
  \mathcal{L}_{\mathrm{DistilBERT}} = - \mathrm{cos\_sim}(h^\tau_{\Ll,t}, W^{\mathrm{map}} h^\sigma_{\Lp,t})
\end{equation}
where $\mathrm{cos\_sim}(\cdot)$ is the vector cosine similarity.


\section{The \vocab~ Dataset}
\label{appendix:vocab_dataset}
We took the list of 479k words from \cite{500k_words} and used Claude 3.5 Sonnet to produce definitions
and important facts about them using the prompt in Listing~\ref{lst:vocab_prompt}.
Listing~\ref{lst:vocab_examples} shows some examples from this dataset.

\lstset{
  basicstyle=\small\ttfamily,    
  columns=fullflexible,    
  breaklines=true,         
  frame=single,            
}
\begin{lstlisting}[caption={Prompt used to produce the \texttt{vocab} dataset.}, label=lst:vocab_prompt, float]  
Please provide ALL the definitions and, if available, important facts about the following terms:  
  
%s  
  
Provide these definitions/facts as a JSON document, formatted as:
{
 "word": ["definition or fact 1", "definition or fact 2", "definition or fact 3", "definition or fact 4", ...]
}  
  
If the word has multiple definitions or facts attached to it, there should be a string for each definition or fact. The list of definitions should be complete.  
  
The definition or fact should contain the term in question, so it can be read stand-alone. For example, the output expected for the terms "subfigure", "linking" and "Pinckney" is:

{  
 "subfigure": [
    "A subfigure is a secondary or smaller figure within a larger figure or diagram, often used in academic and scientific publications."
 ],
 "linking": [
    "Linking is the process of connecting or joining two or more things together.", 
    "In computing, linking is the process of combining various pieces of code and data into a single executable program."
 ],  
 "Pinckney": [
    "Pinckney refers to Charles Cotesworth Pinckney, an American statesman and diplomat.", 
    "Pinckney refers to place names or other historical figures with the surname Pinckney.", 
    "Pinckney's Treaty, also known as the Treaty of San Lorenzo or the Treaty of Madrid, was a diplomatic agreement signed on October 27, 1795, between the United States and Spain. The treaty defined the border between the United States and Spanish Florida at 31 N latitude."
 ]  
}  
  
There should not be any reference between any definition of or facts about a term, so they should NEVER contain the words such as 'can also refer to' or 'also refers to'.
  
Provide your output as a JSON object, and nothing else than the JSON. Do not include any additional descriptions or introductory text.
\end{lstlisting}

\lstset{
  basicstyle=\small\ttfamily,    
  breaklines=true,         
  frame=single,            
}
\begin{lstlisting}[caption={Examples from the \texttt{vocab} dataset.}, label=lst:vocab_examples, float]
{  
  "1080": [  
    "1080 is a natural number following 1079 and preceding 1081.",  
    "1080 refers to a video display resolution of 1920x1080 pixels, also known as Full HD.",  
    "1080 is the chemical compound sodium fluoroacetate, used as a pesticide.",  
    "In mathematics, 1080 is the sum of four consecutive primes (263 + 269 + 271 + 277)."  
  ],  
  "Abipon": [  
    "Abipon refers to an indigenous people who historically inhabited the Gran Chaco region of Argentina.",  
    "The Abipon were known for their horsemanship and warrior culture.",  
    "Abipon is an extinct language once spoken by the Abipon people."  
  ],  
  "abaxial": [  
    "Abaxial in botany refers to the surface of a leaf or other organ facing away from the axis or stem.",  
    "Abaxial is the opposite of adaxial in plant anatomy."  
  ]  
}
\end{lstlisting}

\section{Geometric Comparison Example}
\label{appendix:geometric_comparison_examples}

Table~\ref{tab:sample_sentences} reports the sentences used to produce Figure~\ref{fig:tsne_visualization}.

\section{Scoring Examples}
\label{appendix:scoring_examples}

Table~\ref{tab:sample_scores} shows a comparison of top 10 documents as retrieved by the reference teacher model \snowflakem,
and compared to the results from \adeptir{} run in asymmetric and standard modes.

The documents are retrieved from 100k documents from the CC-News dataset samples of 2024.
The query is ``best marvel movie''.
As can be seen, when run in asymmetric mode there is an 8/10 overlap with the teacher's retrieved documents, while in
standard mode the overlap is 9/10. The top retrieval result, which has a score substantially higher than the rest, is the same
for all three methods.

In Table~\ref{tab:top_10_approximation_errors} we report failure and success modes, in terms of highest and lowest
approximation errors $\| y_i - \yhat_i \|_2$, for both queries and documents. The dataset used for this is the collection of
queries and documents from NanoBEIR.
Texts that involve technical terms (``matrix factorization'', ``dc-dc multilevel boost converter'', ``markov random fields''),
or other languages (including programming languages) appear to be most challenging.
We note that our training datasets do not contain these types of texts, which explains the performance gap and highlights
a potential venue for further improvement.

\begin{table*}[hbtp]
  \caption{Sample sentences from different domains used for the embedding visualization.}
  \label{tab:sample_sentences}
  \small
  \setlength{\tabcolsep}{4pt}
  \centering
  \begin{tabular}{p{0.50\textwidth}|p{0.50\textwidth}}
    \toprule
    \multicolumn{1}{c}{\textbf{Weather}} & \multicolumn{1}{c}{\textbf{Technology}} \\
    \midrule
    The sun peeked through the clouds after a drizzly morning. & The new smartphone features a foldable display and 5G co... \\
    A gentle breeze rustled the leaves as we walked along th... & Quantum computing promises to solve problems beyond clas... \\
    Heavy rains caused flooding in several low-lying neighbo... & Blockchain technology is being explored for secure votin... \\
    It was so hot that even the birds sought shade under the... & Virtual reality headsets are becoming more affordable an... \\
    By midnight, the temperature had dropped below freezing. & The rise of electric vehicles is reshaping the automotiv... \\
    Thunderstorms lit up the sky with flashes of lightning. & Cloud computing allows businesses to scale resources dyn... \\
    A thick fog settled over the city streets at dawn. & Augmented reality applications are transforming retail e... \\
    The air smelled of ozone after the sudden hailstorm. & The Internet of Things connects everyday devices to the... \\
    I watched the snowflakes drift silently onto the ground. & Cybersecurity threats are evolving, requiring constant v... \\
    A double rainbow appeared after the rain shower. & 3D printing is enabling rapid prototyping and custom man... \\
    The humidity soared to uncomfortable levels by midday. & Edge computing reduces latency by processing data close... \\
    Dust devils formed in the dry desert plains. & Biometric authentication methods are enhancing security ... \\
    The barometer readings indicated an approaching front. & Wearable technology is tracking health metrics in real-t... \\
    A sudden gust of wind knocked over the garden chairs. & Artificial intelligence is being used to create realisti... \\
    Light drizzle turned into a torrential downpour within m... & \\
    \midrule
    \multicolumn{1}{c}{\textbf{Cooking}} & \multicolumn{1}{c}{\textbf{Sports}} \\
    \midrule
    Preheat the oven to 375°F before you start mixing the ba... & He dribbled past two defenders and sank a three-pointer ... \\
    She finely chopped the garlic and sautéed it in two tabl... & The marathon runner kept a steady pace despite the swelt... \\
    A pinch of saffron adds a beautiful color and aroma to t... & Their home team clinched the championship with a last-mi... \\
    If the soup is too salty, add a peeled potato to absorb ... & NASCAR fans cheered as the cars roared around the oval t... \\
    Let the bread dough rise for at least an hour in a warm,... & She landed a perfect triple axel at the figure skating c... \\
    Marinate the chicken overnight in a blend of citrus and ... & The cyclist pedaled up the steep hill in record time. \\
    Use a cast-iron skillet to sear the steak on high heat. & He pitched a no-hitter during the high school baseball g... \\
    Whisk the egg whites until they form stiff peaks. & The quarterback threw a touchdown pass under heavy press... \\
    Fold in the chocolate chips gently to keep the batter ai... & They scored a hat-trick in the hockey final. \\
    Brush the pastry with an egg wash for a golden finish. & The boxer delivered a swift uppercut in the final round. \\
    Slow-roast the pork shoulder until it falls off the bone. & Fans erupted when the underdog scored the winning goal. \\
    Garnish the salad with toasted nuts and fresh herbs. & The swimmer broke the national record in the 200m freest... \\
    Deglaze the pan with white wine for a rich sauce. & The gymnast executed a flawless routine on the balance b... \\
    Simmer the curry paste until the aroma intensifies. & The rugby team celebrated their victory with a tradition... \\
    Let the risotto rest before serving to thicken slightly. & \\
    \midrule
    \multicolumn{1}{c}{\textbf{Finance}} & \multicolumn{1}{c}{\textbf{Music}} \\
    \midrule
    The stock market rallied after positive earnings reports. & The symphony orchestra played a hauntingly beautiful mel... \\
    Investors are closely watching interest rate changes by ... & She strummed her guitar softly, filling the room with a ... \\
    Cryptocurrency prices have been extremely volatile this ... & The DJ mixed tracks seamlessly, keeping the crowd dancin... \\
    Diversification is key to managing investment risk effec... & His voice soared during the high notes of the ballad. \\
    Inflation rates have reached a 40-year high, impacting c... & The band played an acoustic set in the intimate coffee s... \\
    Many companies are adopting ESG criteria to attract soci... & Jazz musicians often improvise solos based on the chord ... \\
    The bond market is reacting to geopolitical tensions and... & The opera singer hit the high C with perfect pitch. \\
    Venture capital funding for startups has surged in the t... & The choir harmonized beautifully, filling the church wit... \\
    Exchange-traded funds (ETFs) offer a way to invest in di... & He composed a symphony that was performed at the concert... \\
    The global economy is recovering from the pandemic, but ... & The singer-songwriter wrote heartfelt lyrics about love ... \\
    Central banks are exploring digital currencies to modern... & The rock band headlined the festival, drawing a massive ... \\
    Retail investors are increasingly participating in the s... & Hip-hop artists use rhythm and rhyme to tell powerful st... \\
    Hedge funds are using complex algorithms to gain an edge... & The violinist played a virtuosic solo that left the audi... \\
    Real estate prices have skyrocketed in urban areas due t... & Folk music often reflects the culture and traditions of ... \\
    The startup raised \$10 million in its Series A funding ... & The gospel choir lifted spirits with their uplifting per... \\
    \midrule
    \multicolumn{1}{c}{\textbf{History}} & \multicolumn{1}{c}{\textbf{History}} \\
    \midrule
    The fall of the Berlin Wall in 1989 marked the end of th... & Ancient Egypt's pyramids are a testament to their archit... \\
    Europe's Renaissance period sparked a revival in art and... & The signing of the Declaration of Independence in 1776 e... \\
    The Industrial Revolution transformed economies and soci... & Rome was the center of a vast empire that influenced law... \\
    The discovery of the New World by Christopher Columbus i... & The French Revolution in 1789 led to significant politic... \\
    World War II was a global conflict that reshaped interna... & The fall of the Roman Empire in 476 AD marked the beginn... \\
    The invention of the printing press revolutionized the s... & The Cold War was characterized by political tension betw... \\
    The ancient Silk Road connected East and West through tr... & The signing of the Magna Carta in 1215 established princ... \\
    Exploration during the Age of Discovery expanded Europe... & \\
    \bottomrule
  \end{tabular}
\end{table*}

\begin{table*}
  \caption{Top 10 documents and their scores as retrieved from a collection of 100k news articles by the teacher model, and compared with \adeptir{} run in asymmetric and standard modes.
    A substantial overlap in the documents retrieved and score similarities can be observed.}
  \label{tab:sample_scores}
  \small
  \setlength{\tabcolsep}{4pt}
  \centering
  \begin{tabular}{l|l}
    \textbf{Score} & \textbf{Document}                                                                                       \\ \toprule
    \multicolumn{2}{c}{\textbf{teacher (ground truth)}}                                                                      \\
    \midrule
    0.5904         & The MCU Movie With The Highest Rotten Tomatoes Score - Looper: The MCU Movie With The Highest Rotten... \\
    0.5187         & Kevin Feige: Marvel maestro, box office bellwether: Is Kevin Feige good for the film industry? Depen... \\
    0.4752         & Netflix top 10 movies — here’s the 3 worth watching right now: The Netflix top 10 is a great resourc... \\
    0.4721         & X-Men '97 Created A New Challenge For Marvel's Future: X-Men ’97 has taken the world by storm. The D... \\
    0.4642         & The Best Movies of 2024 So Far: Summer is here! In the old days, that would mean heading to the mult... \\
    0.4632         & 'Indrani' is like a massified version of Marvel movies: Makers: 'Indrani' is like a massified versio... \\
    0.4594         & 5 best movies to watch this weekend on Netflix, Prime Video, Hulu and more: As we head toward summer... \\
    0.4535         & Marvel’s first immersive story for the Apple Vision Pro is the most fun I’ve had on the device: Yes,... \\
    0.4453         & Netflix has just added one of 2023’s very best movies: After its cinema release just last December, ... \\
    0.4420         & Former MCU star just well and truly destroyed rumors of their 'Avengers' comeback, and we're so glad... \\
    \midrule
    \multicolumn{2}{c}{\textbf{\adeptir{} (asym.)}}                                                                           \\
    \midrule
    0.5905         & The MCU Movie With The Highest Rotten Tomatoes Score - Looper: The MCU Movie With The Highest Rotten... \\
    0.5079         & Kevin Feige: Marvel maestro, box office bellwether: Is Kevin Feige good for the film industry? Depen... \\
    0.4729         & X-Men '97 Created A New Challenge For Marvel's Future: X-Men ’97 has taken the world by storm. The D... \\
    0.4621         & Netflix top 10 movies — here’s the 3 worth watching right now: The Netflix top 10 is a great resourc... \\
    0.4618         & Marvel’s first immersive story for the Apple Vision Pro is the most fun I’ve had on the device: Yes,... \\
    0.4548         & 'Indrani' is like a massified version of Marvel movies: Makers: 'Indrani' is like a massified versio... \\
    0.4491         & 5 best movies to watch this weekend on Netflix, Prime Video, Hulu and more: As we head toward summer... \\
    0.4466         & The Best Movies of 2024 So Far: Summer is here! In the old days, that would mean heading to the mult... \\
    0.4394         & Netflix has just added one of 2023’s very best movies: After its cinema release just last December, ... \\
    0.4356         & One of the best hidden gem movies of last year is now streaming on Netflix: Godzilla Minus One tease... \\
    \midrule
    \multicolumn{2}{c}{\textbf{\adeptir}}                                                                                    \\
    \midrule
    0.5925         & The MCU Movie With The Highest Rotten Tomatoes Score - Looper: The MCU Movie With The Highest Rotten... \\
    0.5312         & X-Men '97 Created A New Challenge For Marvel's Future: X-Men ’97 has taken the world by storm. The D... \\
    0.5238         & Kevin Feige: Marvel maestro, box office bellwether: Is Kevin Feige good for the film industry? Depen... \\
    0.4908         & 'Indrani' is like a massified version of Marvel movies: Makers: 'Indrani' is like a massified versio... \\
    0.4823         & Netflix top 10 movies — here’s the 3 worth watching right now: The Netflix top 10 is a great resourc... \\
    0.4681         & One of the best hidden gem movies of last year is now streaming on Netflix: Godzilla Minus One tease... \\
    0.4648         & Marvel’s first immersive story for the Apple Vision Pro is the most fun I’ve had on the device: Yes,... \\
    0.4608         & 5 best movies to watch this weekend on Netflix, Prime Video, Hulu and more: As we head toward summer... \\
    0.4554         & Netflix has just added one of 2023’s very best movies: After its cinema release just last December, ... \\
    0.4505         & The Best Movies of 2024 So Far: Summer is here! In the old days, that would mean heading to the mult... \\
    \bottomrule
  \end{tabular}
\end{table*}

\begin{table*}
  \caption{Top 10 queries and documents with highest and lowest approximation errors.}
  \label{tab:top_10_approximation_errors}
  \small
  \setlength{\tabcolsep}{4pt}
  \centering
  \begin{tabular}{l|l}
    \textbf{Error} & \textbf{Query/Document}                                                                                       \\ \toprule
    \multicolumn{2}{c}{\textbf{Top 10 Queries by Approximation Error (Highest)}}                                                                      \\
    \midrule
    0.5247         & Breaking the Barrier of Transactions: Mining Inter-Transaction Association Rules \\
    0.5195         & What is it like to be an IITian? \\
    0.5062         & Collaborative video reindexing via matrix factorization \\
    0.4932         & Learning to Rank Non-Factoid Answers: Comment Selection in Web Forums \\
    0.4931         & Affordances as a Framework for Robot Control \\
    0.4894         & Photo-Realistic Single Image Super-Resolution Using a Generative Adversarial Network \\
    0.4835         & Novel DC-DC Multilevel Boost Converter \\
    0.4762         & Bank distress in the news: Describing events through deep learning \\
    0.4674         & Petyr Baelish is nicknamed Littlefinger. \\
    0.4673         & Fast Sparse Gaussian Markov Random Fields Learning Based on Cholesky Factorization \\
    \midrule
    \multicolumn{2}{c}{\textbf{Top 10 Documents by Approximation Error (Highest)}}                                                                           \\
    \midrule
    0.6383         & Complete this sentence: Life is boring without…? \\
    0.6262         & Marcos Grigorian (Armenian: Մարկոս Գրիգորեան ; Persian: مارکو گريگوريان‎ ‎ ; December 5, 1925 – Augu... \\
    0.5894         & Big East\textbackslash n1989, 1991, 1994, 1995, 1996, 1997, 1998, 1999, 2000, 2001, 2002, 2005, 2006, 2008, 2009,... \\
    0.5660         & El 7 de octubre de 1941 se fundó la Escuela de Danza, con Ernst Uthoff en la triple labor de directo... \\
    0.5641         & P\textbackslash n\textbackslash n\textbackslash n\textbackslash n\textbackslash n\{\textbackslash displaystyle \{\textbackslash mathfrak \{P\}\}\textbackslash n\textbackslash n5, \textbackslash n\textbackslash n\textbackslash n\textbackslash n\textbackslash nP... \\
    0.5626         & Inductive reasoning allows inferring \textbackslash n\textbackslash n\textbackslash n\textbackslash nb\textbackslash n\textbackslash n\textbackslash n\{\textbackslash displaystyle b\}\textbackslash n\textbackslash n from \textbackslash n\textbackslash n\textbackslash n\textbackslash na... \\
    0.5609         & == Rebuttal == (1) Pro says Chernobyl led to 200,000 deaths. According to recent studies, the actual... \\
    0.5605         & , scanf is vulnerable to format string attacks. Great care should be taken to ensure that the format... \\
    0.5595         & In 1852, Kummer proved that if m and n are nonnegative integers and p is a prime number, then the la... \\
    0.5586         & on the heels of something. Fig. soon after something. There was a rainstorm on the heels of the wind... \\
    \midrule
    \multicolumn{2}{c}{\textbf{Top 10 Queries by Approximation Error (Lowest)}}                                                                    \\
    \midrule
    0.2013         & benefits of oxygen pills \\
    0.2098         & where are the cones in the eye located \\
    0.2150         & how long can i keep cooked italian sausage in refrigerator \\
    0.2165         & Is cell phone radiation safe? \\
    0.2198         & Is obesity a disease? \\
    0.2230         & Foods for Glaucoma \\
    0.2239         & airport scanners \\
    0.2259         & types of ethnic foods list \\
    0.2263         & What are the types of bariatric surgery? \\
    0.2276         & side effects steroid injection \\
    \midrule
    \multicolumn{2}{c}{\textbf{Top 10 Documents by Approximation Error (Lowest)}}                                                                    \\
    \midrule
    0.1764         & How \\
    0.1804         & Legitimacy \\
    0.1809         & Rudd is a surname of English, Scottish, Welsh and Jewish origin. \\
    0.1812         & The Colorado River is one of the principal rivers of the Southwestern United States and northern Mex... \\
    0.1821         & Glacier National Park is a national park located in the U.S. state of Montana , on the Canada -- Uni... \\
    0.1864         & Dillon is an unincorporated community in Botetourt County, Virginia, United States. \\
    0.1884         & A bandmaster is the leader and conductor of a band, usually a military or marching band. \\
    0.1887         & what \\
    0.1887         & what \\
    0.1887         & what \\
    \bottomrule
  \end{tabular}
\end{table*}

\section{MTEB v2 (Eng) Results}
\label{appendix:mteb_v2_scores_adept_mt}

Table~\ref{tab:combined_results} reports the results of the MTEB v2 (English) benchmark on the 41 individual datasets
that constitute it.

Results for \adeptmt{} standard and asymmetric modes are identical except on reranking and retrieval tasks, as these
are the only tasks that have query and document sides.

Highlighted in bold are cases where the distilled model performs better than the teacher. In many of these cases,
these differences are negligible except for Touche2020Retrieval.v3 and StackExchangeClustering.v2. We have manually
checked the former and verified that the results are correct.

\begin{table*}[p]
  \centering
  \caption{MTEB v2 (English) benchmark performance. \adeptmt{} substantially
    outperforms the teacher model on Touche2020Retrieval.v3 and StackExchangeClustering.v2. Highlighted are datasets in which \adeptmt{} outperforms the teacher. (*) These values are identical in standard and asymmetric modes. $^\dagger$\mxbai~ values
    are taken from the online version of the MTEB leaderboard.}
  \label{tab:combined_results}
  \small
  \setlength{\tabcolsep}{5pt}
  \begin{tabular}{lccccc}
    \textbf{Task Name}                              & \rotatebox{45}{\textbf{\adeptmt}} & \rotatebox{45}{\textbf{\adeptmt{} (asym.)}} & \rotatebox{45}{\textbf{mxbai-l-v1}$^\dagger$} & \textbf{Metric}    & \textbf{Task Type}      \\
    \midrule
    AmazonCounterfactualClassification              & 71.27                             & *                                          & 75.07                                         & Accuracy           & Classification          \\
    Banking77Classification                         & 85.07                             & *                                          & 87.80                                         & Accuracy           & Classification          \\
    ImdbClassification                              & 89.05                             & *                                          & 92.83                                         & Accuracy           & Classification          \\
    MTOPDomainClassification                        & 92.73                             & *                                          & 93.95                                         & Accuracy           & Classification          \\
    MassiveIntentClassification                     & 72.35                             & *                                          & 76.22                                         & Accuracy           & Classification          \\
    MassiveScenarioClassification                   & 77.04                             & *                                          & 79.89                                         & Accuracy           & Classification          \\
    ToxicConversationsClassification                & 67.50                             & *                                          & 67.31                                         & Accuracy           & Classification          \\
    \textbf{TweetSentimentExtractionClassification} & \textbf{60.23}                    & *                                          & \textbf{59.70}                                & \textbf{Accuracy}  & \textbf{Classification} \\
    \midrule
    ArXivHierarchicalClusteringP2P                  & 59.50                             & *                                          & 60.11                                         & V Measure          & Clustering              \\
    ArXivHierarchicalClusteringS2S                  & 52.91                             & *                                          & 58.99                                         & V Measure          & Clustering              \\
    BiorxivClusteringP2P.v2                         & 41.76                             & *                                          & 41.87                                         & V Measure          & Clustering              \\
    \textbf{MedrxivClusteringP2P.v2}                & \textbf{37.58}                    & *                                          & \textbf{37.27}                                & \textbf{V Measure} & \textbf{Clustering}     \\
    MedrxivClusteringS2S.v2                         & 34.49                             & *                                          & 34.97                                         & V Measure          & Clustering              \\
    \textbf{StackExchangeClustering.v2}             & \textbf{58.34}                    & *                                          & \textbf{55.08}                                & \textbf{V Measure} & \textbf{Clustering}     \\
    StackExchangeClusteringP2P.v2                   & 40.28                             & *                                          & 40.47                                         & V Measure          & Clustering              \\
    TwentyNewsgroupsClustering.v2                   & 47.29                             & *                                          & 51.10                                         & V Measure          & Clustering              \\
    \midrule
    SprintDuplicateQuestions                        & 96.48                             & *                                          & 96.82                                         & AP                 & PairClassification      \\
    TwitterSemEval2015                              & 71.25                             & *                                          & 78.55                                         & AP                 & PairClassification      \\
    TwitterURLCorpus                                & 85.64                             & *                                          & 86.23                                         & AP                 & PairClassification      \\
    \midrule
    AskUbuntuDupQuestions                           & 61.35                             & 62.03                                      & 63.49                                         & MAP                & Reranking               \\
    MindSmallReranking                              & 32.35                             & 32.53                                      & 32.62                                         & MAP                & Reranking               \\
    \midrule
    ArguAna                                         & 61.64                             & 64.44                                      & 65.47                                         & nDCG@10            & Retrieval               \\
    CQADupstackGamingRetrieval                      & 54.29                             & 54.00                                      & 58.94                                         & nDCG@10            & Retrieval               \\
    CQADupstackUnixRetrieval                        & 36.58                             & 37.25                                      & 41.77                                         & nDCG@10            & Retrieval               \\
    ClimateFEVERHardNegatives                       & 26.79                             & 35.96                                      & 36.23                                         & nDCG@10            & Retrieval               \\
    FEVERHardNegatives                              & 72.49                             & 81.13                                      & 86.54                                         & nDCG@10            & Retrieval               \\
    FiQA2018                                        & 36.27                             & 42.22                                      & 45.27                                         & nDCG@10            & Retrieval               \\
    HotpotQAHardNegatives                           & 62.23                             & 65.07                                      & 72.50                                         & nDCG@10            & Retrieval               \\
    SCIDOCS                                         & 18.09                             & 19.06                                      & 23.10                                         & nDCG@10            & Retrieval               \\
    TRECCOVID                                       & 52.52                             & 73.03                                      & 75.53                                         & nDCG@10            & Retrieval               \\
    \textbf{Touche2020Retrieval.v3}                 & \textbf{51.58}                    & \textbf{45.72}                             & \textbf{48.60}                                & \textbf{nDCG@10}   & \textbf{Retrieval}      \\
    \midrule
    BIOSSES                                         & 84.45                             & *                                          & 86.05                                         & Spearman           & STS                     \\
    SICK-R                                          & 80.87                             & *                                          & 82.78                                         & Spearman           & STS                     \\
    STS12                                           & 77.96                             & *                                          & 79.07                                         & Spearman           & STS                     \\
    STS13                                           & 87.93                             & *                                          & 89.79                                         & Spearman           & STS                     \\
    STS14                                           & 82.40                             & *                                          & 85.22                                         & Spearman           & STS                     \\
    STS15                                           & 88.08                             & *                                          & 89.34                                         & Spearman           & STS                     \\
    STS17                                           & 87.12                             & *                                          & 89.21                                         & Spearman           & STS                     \\
    \textbf{STS22.v2}                               & \textbf{69.12}                    & *                                          & \textbf{69.04}                                & \textbf{Spearman}  & \textbf{STS}            \\
    STSBenchmark                                    & 87.19                             & *                                          & 89.29                                         & Spearman           & STS                     \\
    \midrule
    SummEvalSummarization.v2                        & 30.86                             & *                                          & 32.63                                         & Spearman           & Summarization           \\
    \bottomrule
  \end{tabular}
\end{table*}

\section{Robustness Margins for \adeptmt}
\label{appendix:robustness_adept_mt}

\begin{figure}[hbtp]
  \centering
  \caption{NanoBEIR performance of various checkpoints stored while training \adeptmt. Similar to the discussion in Section~\ref{sec:systems_robustness}, we observe substantial robustness margins for this model.}
  \label{fig:mxbai_robustness}
  \includegraphics[width=1\linewidth]{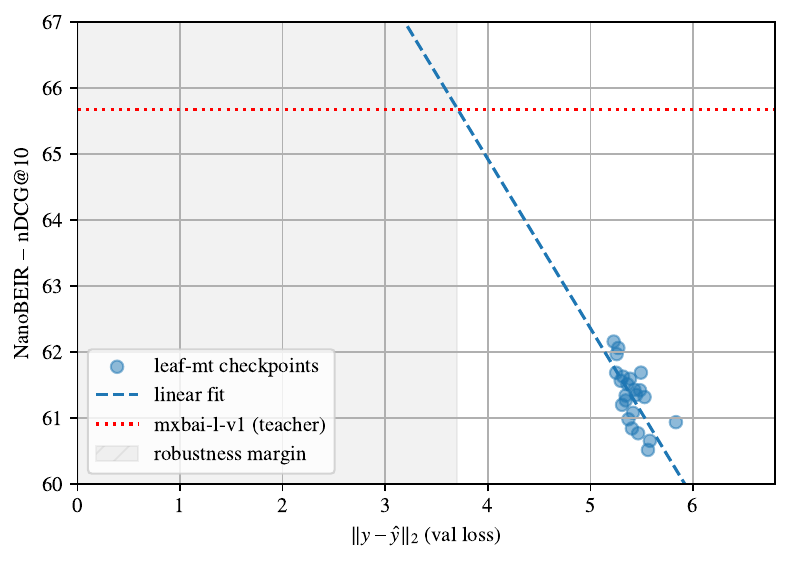}
\end{figure}

Section~\ref{sec:systems_robustness} discusses our observed systems' robustness to
the approximation error introduced by our knowledge distillation scheme.
The robustness margin for \adeptmt{} is shown in Figure~\ref{fig:mxbai_robustness}.
Note that the teacher model for \adeptmt, \mxbai, does not output normalized vectors,
and as such the approximation error $\epsilon$ does not have an upper bound of 2.

\section{Inference Time Performance (CPU)}
\label{appendix:inference_time_speed}

Increasing inference throughput, lowering latencies, and operating on more economical infrastructure are typical core
goals of model distillation activities.

Table~\ref{tab:cpu_timing} reports our results for these key metrics of \adeptir{} and \adeptmt{} when compared to their corresponding teachers.

The experiments are conducted on an AWS EC2 \verb|i3.large| instance, which is a commonly used CPU-only VM for database and search workloads.
These instances are equipped with 2 vCPUs and 15.25GB of RAM. Inference is performed on the ONNX Runtime.

\begin{table*}[ht]
  \caption{Performance comparison across different models. Max $N$ is the largest batch size that can be computed in 100ms or less; ``-'' indicates that no batch size can be computed within this time constraint.}
  \label{tab:cpu_timing}
  \small
  \setlength{\tabcolsep}{3pt}
  \centering
  \begin{tabular}{llccccccc}
                   & \multicolumn{4}{c}{\textbf{docs}} & \multicolumn{4}{c}{\textbf{queries}}                                                                                                                                                           \\
    \cmidrule(lr){2-5} \cmidrule(lr){6-9}
    \textbf{model} & \textbf{docs/s}                   & \textbf{speedup}                     & \textbf{min latency}                & \textbf{max $N$} & \textbf{queries/s} & \textbf{speedup} & \textbf{min latency}                & \textbf{max $N$} \\
    \midrule
    \snowflakem    & $3.2 \pm 0.8$                     & $1\times$                            & $191 \pm 52.6~\text{ms}$  & -                & $16.7 \pm 1.4$     & $1\times$        & $73.2 \pm 5.8~\text{ms}$  & 1                \\
    \adeptir       & $21.9 \pm 5.5$                    & $6.5\times$                          & $28.2 \pm 9.69~\text{ms}$ & 2                & $121.3 \pm 14.0$   & $7.3\times$      & $11.4 \pm 1.38~\text{ms}$ & 8                \\
    \midrule
    \mxbai         & $0.9 \pm 0.2$                     & $1\times$                            & $834 \pm 298~\text{ms}$   & -                & $4.9 \pm 0.4$      & $1\times$        & $236 \pm 18.5~\text{ms}$  & -                \\
    \adeptmt       & $21.4 \pm 5.8$                    & $24.4\times$                         & $39.2 \pm 10.2~\text{ms}$ & 2                & $117.0 \pm 15.9$   & $23.7\times$     & $11.5 \pm 1.2~\text{ms}$  & 8                \\
    \bottomrule
  \end{tabular}
\end{table*}

\begin{figure*}[hbtp]
  \centering
  \caption{\adeptir{} wall clock inference times on an \texttt{i3.large} instance for queries and documents, as a function of batch size. \leaf{} models allow for larger batch sizes while remaining under the 0.1 seconds threshold for queries.}
  \label{fig:cpu_timing}
  \includegraphics[width=\linewidth]{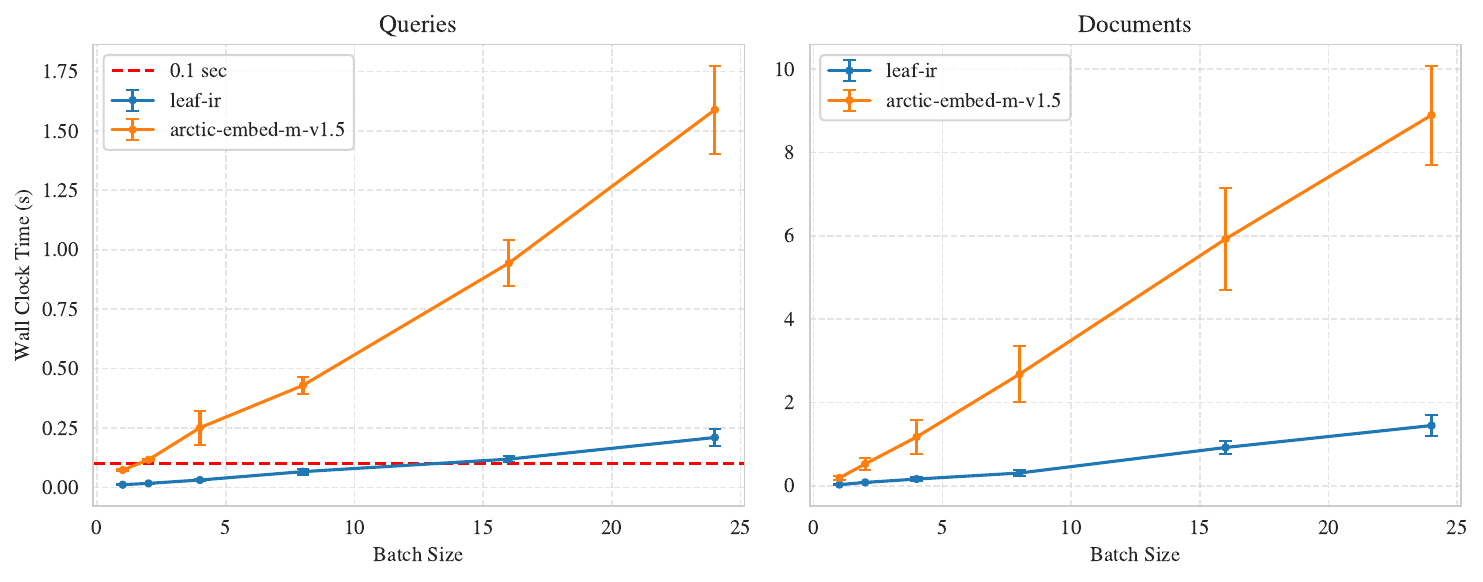}
\end{figure*}

To obtain these measurements, we first select the following set of batch sizes: 1, 2, 4, 8, 16 and 24.
For each batch size, we randomly sample a corresponding number of queries and documents from MSMARCO.
Queries are generally much shorter than documents in this dataset; they reflect the typical user queries of an online search engine.
We use Python's \verb|timeit| package to measure the wall clock time required to complete the inference of a batch 7 times.
As shown in Figure~\ref{fig:cpu_timing}, inference time scales approximately linearly with batch size.
We then divide this time by the batch size to derive the throughput metrics (docs/sec and queries/sec) reported in the Table.
Mean and standard deviation for these throughput figures are computed across all the samples for all the batch sizes.

Our results show that for \adeptir, the 4.7$\times$ parameter reduction leads to a $6.5\times$ throughput increase
when used to encode documents, and a $7.3\times$ throughput increase for queries. For \adeptmt, on the other hand, the
$14.6\times$ parameter reduction leads to a $24.4\times$ throughput increase for docs, and $23.7\times$ throughput increase for queries.

Table~\ref{tab:cpu_timing} also reports the minimum latency, i.e., the shortest amount of time a user would need to wait for
inference to complete. In all of our experiments, this occurs at a batch size of 1.
Industry research and usability studies, including \cite{nielsen_throughput}, commonly highlight 0.1 seconds as the
responsiveness threshold for cognitive perception in user-facing applications.
In the Table we thus also report the largest batch size that can be processed while remaining below this threshold. We report
these metrics both for queries as well as documents, although this measurement is more critical for queries. We also show
this as a red dashed line in Figure~\ref{fig:cpu_timing}. A dash ``\verb|-|'' in the table indicates that no batch size exists that can be computed
under 0.1 seconds.


As shown, \adept{} models have substantially lower minimum latencies than their counterpart teachers. They are also
the only models that can process a batch size within the 0.1 seconds threshold on our test hardware configuration,
with the exception of \snowflakem~ when processing queries.

\section{MRL Results}
\label{appendix:mrl_transfer_mt}

\begin{figure*}[htbp]
  \centering
  \includegraphics[width=\linewidth]{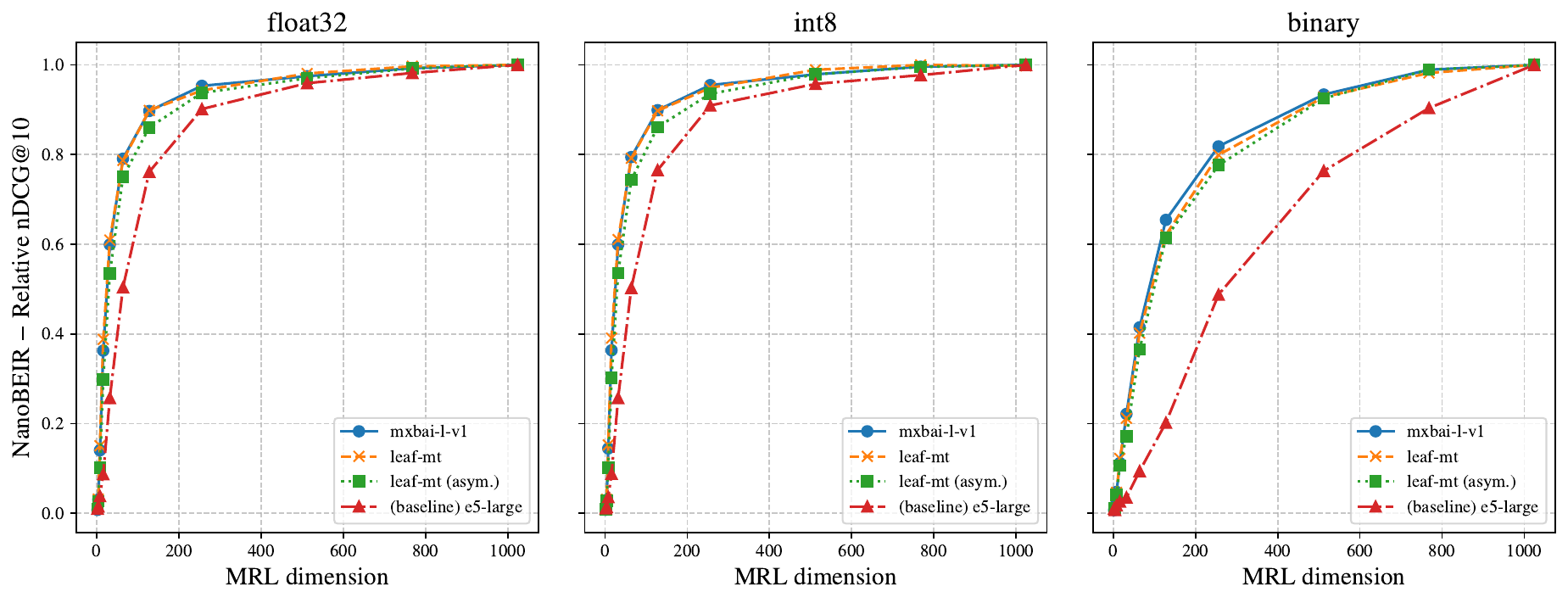}
  \caption{Relative MRL performance of \adeptmt.}
  \label{fig:mrl_transfer}
\end{figure*}

\begin{figure*}[htbp]
  \centering
  \begin{subfigure}[b]{\linewidth}
    \centering
    \includegraphics[width=\linewidth]{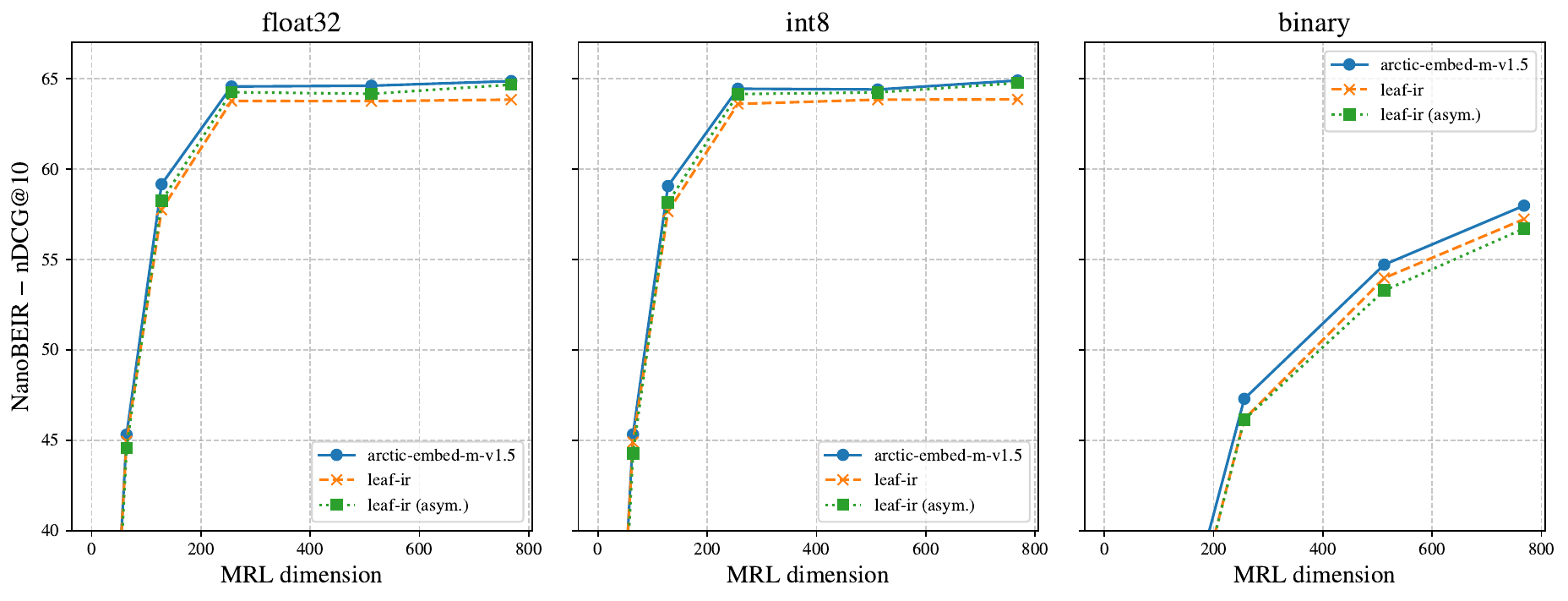}
    \caption{\adeptir}
    \label{fig:mrl_ir}
  \end{subfigure}

  \vspace{1em}

  \begin{subfigure}[b]{\linewidth}
    \centering
    \includegraphics[width=\linewidth]{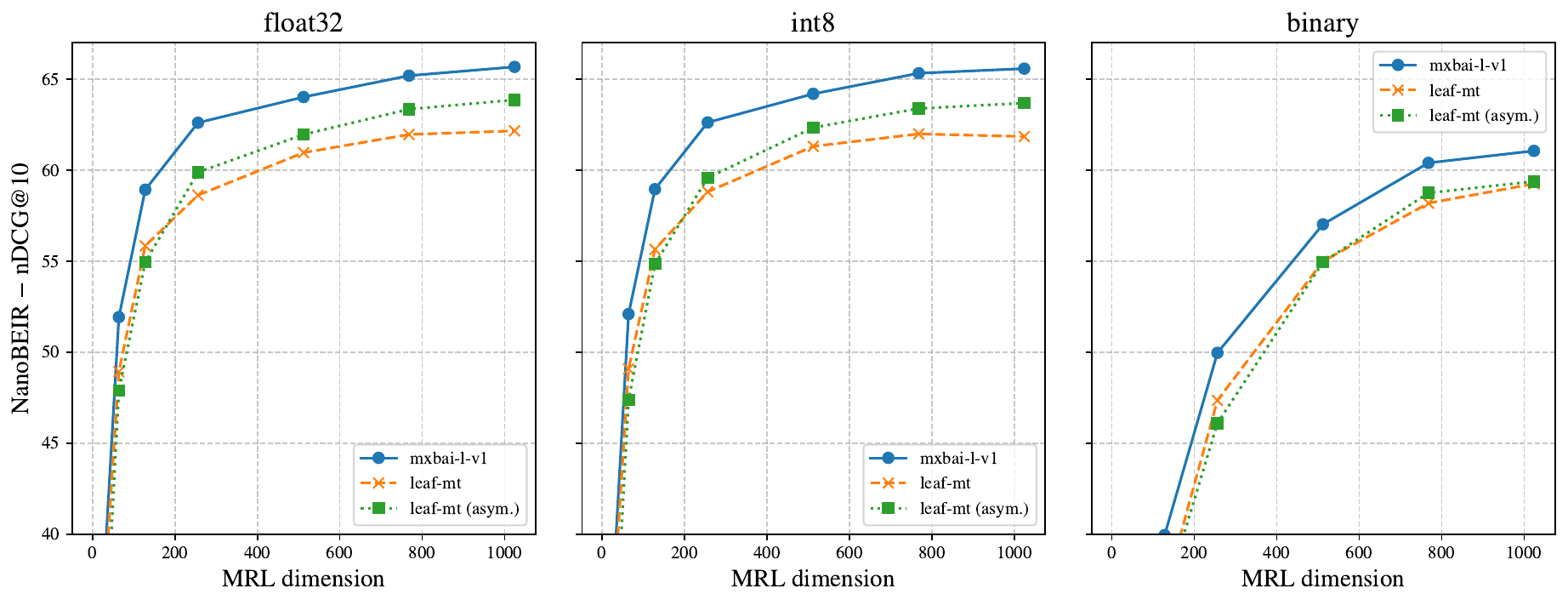}
    \caption{\adeptmt{}}
    \label{fig:mrl_mt}
  \end{subfigure}
  \caption{Absolute MRL performance of \adeptir{} and \adeptmt.}
  \label{fig:absolute_mrl_performance}
\end{figure*}

Figure~\ref{fig:mrl_transfer} shows that, similarly to \adeptir, \adeptmt{} also acquires MRL and quantization properties
from its teacher, while Figure~\ref{fig:absolute_mrl_performance} reports the absolute MRL performance curves for these two models.

\end{document}